\documentclass[a4paper,fleqn]{cas-sc}

\usepackage[numbers]{natbib}

\usepackage{graphicx}
\usepackage{bm}
\usepackage{physics}
\usepackage[utf8]{inputenc}
\usepackage[T1]{fontenc}
\usepackage{etoolbox}
\usepackage{subcaption}
\usepackage{color}
\usepackage[whole]{bxcjkjatype}
\usepackage{siunitx}
\usepackage{comment}
\usepackage{soul}

\def\tsc#1{\csdef{#1}{\textsc{\lowercase{#1}}\xspace}}

\tsc{WGM}
\tsc{QE}

\begin{document}
\let\WriteBookmarks\relax
\def\floatpagepagefraction{1}
\def\textpagefraction{.001}

\shorttitle{}    

\shortauthors{}  

\title [mode = title]{A generalized energy-consistent finite difference scheme for 10-moment magnetohydrodynamics}  



%

\author[1]{Keita Akutagawa}[orcid=0009-0005-4643-1660]



\ead{akutagawa@eps.s.u-tokyo.ac.jp}


\credit{Conceptualization, Investigation, Methodology, Software, Validation, Visualization, Writing - original draft}

\affiliation[1]{organization={Department of Earth and Planetary Science, Graduate School of Science, The University of Tokyo},
            addressline={7-3-1 Hongo, Bunkyo-ku}, 
            city={Tokyo},
            postcode={113-8654}, 
            country={Japan}}

\author[1]{Shinsuke Imada}[orcid=0000-0001-7891-3916]

\credit{Funding acquisition, Writing - review, Supervision}

\author[1]{Munehito Shoda}[orcid=0000-0002-7136-8190]

\credit{Funding acquisition, Investigation, Methodology, Validation, Writing - review and editing, Supervision}





\cortext[1]{Corresponding author}



\begin{abstract}
Pressure anisotropy and off-diagonal pressure stresses are ubiquitous and play important roles in collisionless/weakly collisional plasmas. The Chew–Goldberger–Low (CGL) MHD model is often used; however, it can lose hyperbolicity when the pressure anisotropy or plasma beta becomes large, making it hard to develop approximate Riemann solvers. An alternative approach is to use the 10-moment MHD equations, but their eigenmode analysis is also difficult, which similarly hinders the development of less-diffusive Riemann solvers. This paper presents a new energy-consistent finite difference scheme for 10-moment MHD designed to operate over a broad range of plasma beta. The proposed scheme extends the one-fluid 10-moment MHD model of \citet{hirabayashi2016} using the energy-consistent finite-difference approach developed for ideal MHD by \citet{iijima2021}. Nonlinear filtering is applied to all six independent components of the pressure tensor, and the kinetic and magnetic energies dissipated by the filtering are explicitly transferred to the diagonal pressure components under an equipartition assumption to maintain consistency with the total energy balance. The proposed scheme is validated against seven test problems in the isotropic limit, the gyrotropic limit, and without isotropization/gyrotropization. The results demonstrate the expected spatial convergence and total energy behavior, reproduce the linear growth of the parametric decay and firehose instabilities, and yield pressure tensor structures qualitatively consistent with theoretical expectations and previous simulations in test problems of the blast-wave problem and magnetic reconnection, spanning plasma beta values from $10^{-10}$ to $10^{10}$. The proposed scheme provides a promising framework for large-scale simulations of collisionless plasmas across widely separated plasma beta regimes and opens a path toward applications such as solar wind turbulence and plasmoid-mediated reconnection.
\end{abstract}


\begin{highlights}
\item An energy-consistent finite difference scheme is proposed for 10-moment MHD. 
\item The scheme robustly treats pressure anisotropy and off-diagonal pressure stresses. 
\item The scheme remains stable across plasma beta values from $10^{-10}$ to $10^{10}$. 
\end{highlights}


\maketitle


\section{Introduction}\label{chap1:introduction}

Magnetohydrodynamics (MHD) model is widely used to understand fluid-scale plasma dynamics. It can treat large-scale dynamics, however it does not include kinetic effects, which play important roles in plasma phenomena such as the dynamics of the diffusion region in magnetic reconnection and the transition region of collisionless shocks. The particle-in-cell (PIC) model, in contrast, is often used to capture plasma kinetics, but it cannot treat large-scale dynamics owing to its high computational cost. The hybrid model, in which ions are treated as particles and electrons as a fluid, can capture larger systems than PIC; however, it does not include electron-scale kinetic effects and remains computationally too expensive to treat large-scale phenomena (larger than $10^{3-4}$ ion inertial length systems). Developing numerical models and schemes that incorporate essential kinetic effects while remaining applicable to large-scale systems is therefore an important challenge in space plasma physics, particularly when there is a vast separation between fluid and kinetic scales, as in solar flares, where the scale separation can reach a factor of $\sim 10^{7}$--$10^{8}$.

To date, various extensions of conventional MHD have been developed to incorporate kinetic effects \cite{hakim2008}. One of the best-known examples is the Chew--Goldberger--Low (CGL) model, which accounts for pressure anisotropy by treating the pressures parallel and perpendicular to the magnetic field independently \citep{chew1956}. The CGL-MHD model has been widely applied to studies of the Earth's magnetosphere \citep{hesse1992, denton1996, erkaev1999}, magnetic reconnection \citep{hirabayashi2013}, and turbulence \citep{santos-lima2014}. A key numerical difficulty in the CGL-MHD model lies in the treatment of regions where the magnetic field approaches zero, since the magnetic field strength $|\bm{B}|$ appears in the denominators. Moreover, as the pressure anisotropy becomes sufficiently large, the CGL-MHD system loses hyperbolicity because some of its characteristic eigenvalues become complex \cite{kato1966}. Even recently, new schemes have been developed \citep{luo2023, bhoriya2024, singh2025} to achieve more stable numerical simulations that account for anisotropic effects; a history of anisotropic MHD schemes is reviewed by \citet{bhoriya2024}.

Another approach is to use the 10-moment MHD equations, which can treat all six components of the pressure tensor and thereby encompass both ideal and anisotropic MHD \citep{hakim2008, wang2015, ng2020, walters2024, samuel2025arXiv}. Because the pressure tensor is treated directly, this approach does not require special treatment of magnetic null point
, where the CGL closure becomes invalid. Moreover, it can capture the effect of the off-diagonal pressure tensor components, which are important for momentum transport (effective viscosity). \citet{hirabayashi2016} derived a one-fluid 10-moment MHD model from the Vlasov-Maxwell system and developed a corresponding numerical scheme based on an HLL approximate Riemann solver. They performed several test simulations, such as a one-dimensional Riemann problem for a reconnection layer, and found that the proposed model and scheme correctly treat the isotropic and anisotropic limits, as well as cases without isotropization/gyrotropization. One limitation of this approach is that the use of an approximate Riemann solver causes the amount and partition of numerical heating to be determined implicitly by the numerical scheme. This can be particularly problematic when anisotropic heating plays an important role in the plasma dynamics. Another limitation is that developing less diffusive Riemann solvers is difficult because the analytical eigenmodes of the 10-moment system are difficult to obtain. These limitations motivate the development of an alternative approach that does not rely on approximate Riemann solvers.

Recently, a new finite-difference-based numerical scheme has been developed that remains stable even in extremely low beta systems ($\beta \sim 10^{-10}$) \citep{iijima2021}. Instead of solving the total energy equation, the scheme directly evolves the internal energy equation and explicitly accounts for the numerical dissipation introduced by nonlinear filtering to suppress the numerical oscillation. The kinetic and magnetic energies dissipated by the artificial diffusivity are transferred to the internal energy equation to preserve the total energy conservation. Although the scheme was originally formulated for conventional MHD, its finite-difference formulation does not rely on the explicit form of the characteristic eigenmodes, suggesting that it may be readily extended to more general MHD models. In particular, unlike the Riemann solvers commonly used in previous 10-moment simulations, this approach allows numerical heating to be treated explicitly. This makes it possible to control how the dissipated energy is redistributed among the pressure components, thereby directly addressing the difficulty of controlling the direction of numerical heating.

This study aims to develop a new, stable 10-moment MHD solver. Our strategy is to solve the 10-moment MHD model of \citet{hirabayashi2016} by extending the scheme of \citet{iijima2021}. The advantage of 
this approach is that it can treat anisotropic effects in both high-beta systems, where the CGL closure becomes numerically unstable, and low-beta systems, where plasma dynamics and the associated heating are inherently anisotropic. 
Moreover, because a finite difference method is used, there is no need to derive the eigenmodes of the 10-moment MHD system analytically. This avoids the remaining challenge of the 10-moment MHD approach noted above, namely the partition of numerical heating and the difficulty of developing a less diffusive Riemann solver.

The remainder of this paper is organized as follows. Section \ref{chap2:review_of_10-moment_MHD} reviews the 10-moment MHD equations derived by \citet{hirabayashi2016}. Section \ref{chap3:review_of_iijima_scheme} reviews the robust ideal MHD scheme proposed by \citet{iijima2021}. Section \ref{chap4:extension_to_10-moment_MHD} presents our new scheme for the 10-moment MHD system. Section \ref{chap5:test_problems} shows the results of the test simulations. Finally, Section \ref{chap6:conclusion} summarizes our scheme and discusses some future applications. In what follows, repeated Greek indices $\alpha, \beta, \gamma, \dots$ imply summation over spatial coordinates according to the Einstein summation convention: $P_\alpha Q_\alpha := \sum_\alpha P_\alpha Q_\alpha$. Latin indices $i, j, k, \dots$ are used as free spatial indices. The Levi--Civita symbol is denoted by $\epsilon_{\alpha\beta\gamma}$, with the cross product expressed as $(\bm{P} \times \bm{Q})_i = \epsilon_{i\alpha\beta} P_\alpha Q_\beta$. We also use the identity $\epsilon_{\alpha ij}\epsilon_{\alpha kl} = \delta_{ik}\delta_{jl} - \delta_{il}\delta_{jk}$.


\section{Derivation of 10-moment MHD equations}\label{chap2:review_of_10-moment_MHD}

We first briefly review the derivation of the non-relativistic 10-moment MHD equations from the Vlasov--Maxwell equations following \citet{hirabayashi2016}. We set the vacuum permeability to $\mu_0 = 1$. The Vlasov equation for species $s$ is as follows: 
\begin{align}
    \frac{\partial f_s}{\partial t} + v_{s, \alpha} \partial_\alpha f + \frac{q_s}{m_s} (E_\alpha + \epsilon_{\alpha \beta \gamma} v_{s, \beta} B_\gamma) \partial_{v_\alpha} f_s = 0, 
    \label{chap2:eq:vlasov}
\end{align}
where $f_s$ is the distribution function, $q_s$ the charge density, and $m_s$ the mass of species $s$. $\bm{E}$ and $\bm{B}$ are the electric and magnetic fields. 

Here we define moment variables $n_s, \bm{V}_s, \bm{P}_s, \bm{Q}_s$ as follows: $\int f_s \dd{\bm{v}_s} = n_s, \int f_s \bm{v}_s \dd{\bm{v}_s} = n_s \bm{V}_s, \int f_s \bm{v}_s \bm{v}_s \dd{\bm{v}_s} = n_s \bm{V}_s \bm{V}_s + \bm{P}_s / m_s, \int f_s \bm{v}_s \bm{v}_s \bm{v}_s \dd{\bm{v}_s} = n_s \bm{V}_s \bm{V}_s \bm{V}_s + (\bm{V}_s \bm{P}_s)^{\mathscr{S}} / m_s + \bm{Q}_s / m_s$ where $(\bm{V}_s \bm{P}_s)^{\mathscr{S}}_{ijk} := V_i P_{jk} + V_j P_{ki} + V_k P_{ij}$. Taking the zeroth, first and second moment in Eq.\eqref{chap2:eq:vlasov} for the velocity $\bm{v}_s$, summing up all species, and assuming an ion-electron system ($s = \rm{i, e}$) with $m_{\rm i} / m_{\rm e} \gg 1$, we derive one-fluid moment equation as follows: 
\begin{align}
    &\partial_t \rho + \partial_\alpha (\rho V_\alpha) = 0, \label{chap2:eq:zeroth_moment}\\ 
    &\partial_t (\rho V_i) + \partial_\alpha (\rho V_i V_\alpha + P_{i \alpha})
    = \epsilon_{i \alpha \beta} J_\alpha B_\beta, \label{chap2:eq:first_moment} \\
    &\partial_t (\rho V_i V_j + P_{ij}) + \partial_\alpha (\rho V_i V_j V_\alpha + V_i P_{j \alpha} + V_j P_{\alpha i} + V_\alpha P_{ij} + Q_{ij\alpha}) \label{chap2:eq:second_moment_init}
    = M_{ij} + M_{ji}, \nonumber\\
    &M_{ij} := \left( \bm{J} \bm{E} + \sum_s q_s n_s \bm{V}_s \bm{V}_s \times \bm{B} + \sum_s \frac{q_s}{m_s} \bm{P}_s \times \bm{B} \right)_{ij}. 
\end{align} 
where $\rho = \sum_s m_s n_s$, $\rho \bm{V} = \sum_s m_s n_s \bm{V}_s, \rho \bm{V} \bm{V} + \bm{P} = \sum_s (m_s n_s \bm{V}_s \bm{V}_s + \bm{P}_s), \rho \bm{V} \bm{V} \bm{V} + (\bm{V} \bm{P})^{\mathscr{S}} + \bm{Q} = \sum_s (m_s n_s \bm{V}_s \bm{V}_s \bm{V}_s + (\bm{V}_s \bm{P}_s)^{\mathscr{S}} + \bm{Q}_s)$, and $\bm{J} = \sum_s q_s n_s \bm{V}_s$ is the current. The third term of $M_{ij}$ corresponds to the rotation of the pressure tensor $P_{ij}$ due to the gyration motion of particles. We do not treat this term directly because resolving the gyration period is computationally expensive; instead, we introduce a model that mimics the gyration effect, as explained later.

The second term of $M_{ij}$ can be written as: 
\begin{align}
    q n (\bm{V}_{\rm i} \bm{V}_{\rm i} - \bm{V}_{\rm e} \bm{V}_{\rm e}) 
    \sim q n \left( \bm{V} \bm{V} - \left( \bm{V} - \frac{\bm{J}}{q n} \right) \left( \bm{V} - \frac{\bm{J}}{q n} \right) \right) 
    = \left( \bm{V} \bm{J} + \bm{J} \bm{V} - \frac{\bm{JJ}}{q n} \right), 
\end{align}
where $q_{\rm i} = |q_{\rm e}| = q$, and we assume quasi-neutrality $n_{\rm i} = n_{\rm e}$ and only electrons contribute to current. Then, we get $M_{ij}$ as follows: 
\begin{align}
    M_{ij} = \left( \bm{J} \bm{E} + \left( \bm{V} \bm{J} + \bm{J} \bm{V} - \frac{\bm{JJ}}{q n} \right) \times \bm{B} \right)_{ij} 
    = \left( \bm{V} (\bm{J} \times \bm{B}) + \bm{J} \left( \bm{E} + \bm{V} \times \bm{B} - \frac{\bm{J} \times \bm{B}}{qn} \right) \right)_{ij}. 
\end{align}
When we focus on phenomena whose spatial and temporal scales are much larger than the inertial or gyration scales, we can ignore the Hall electric field and adopt the ideal Ohm's law $\bm{E} = -\bm{V} \times \bm{B}$, and $M_{ij}$ is rewritten as follows:
\begin{align}
    M_{ij} = (\bm{V} (\bm{J} \times \bm{B}))_{ij} = V_i \epsilon_{j \alpha \beta} J_\alpha B_\beta. 
\end{align}
In this paper, we set the heat flux $\bm{Q} = \bm{0}$. This is justified when the spatial and temporal scales of interest are much larger than those associated with phenomena such as temperature relaxation or Landau damping. Then, Eq.~\eqref{chap2:eq:second_moment_init} becomes 
\begin{align}
    \partial_t (\rho V_i V_j + P_{ij}) + \partial_\alpha (\rho V_i V_j V_\alpha + V_i P_{j \alpha} + V_j P_{\alpha i} + V_\alpha P_{ij}) 
    = V_i \epsilon_{j \alpha \beta} J_\alpha B_\beta + V_j \epsilon_{i \alpha \beta} J_\alpha B_\beta. 
    \label{chap2:eq:second_moment}
\end{align}
Using $\bm{J} = \nabla \times \bm{B}$ (i.e., neglecting the displacement current), the RHS is rewritten as follows: 
\begin{align}
    {\rm{RHS}} = B_\alpha V_i \partial_\alpha B_j + B_\alpha V_j \partial_\alpha B_i - B_\alpha V_i \partial_j B_\alpha - B_\alpha V_j \partial_i B_\alpha
\end{align}

From the induction equation, we obtain the equation for the magnetic tensor as follows: 
\begin{align}
    \partial_t (B_i B_j) + \partial_\alpha (\epsilon_{i \alpha \beta} B_j E_\beta + \epsilon_{j \alpha \beta} B_i E_\beta) = \epsilon_{i \alpha \beta} (\partial_\alpha B_j) E_\beta + \epsilon_{j \alpha \beta} (\partial_\alpha B_i) E_\beta. 
    \label{chap2:eq:magnetic_tensor}
\end{align}
In the ideal MHD case $\bm{E} = -\bm{V} \times \bm{B}$, and the RHS is rewritten as follows: 
\begin{align}
    {\rm{RHS}} = -B_\alpha V_i \partial_\alpha B_j - B_\alpha V_j \partial_\alpha B_i + B_i V_\alpha \partial_\alpha B_j + B_j V_\alpha \partial_\alpha B_i. 
\end{align}
Finally, adding Eq.~\eqref{chap2:eq:second_moment} and ~\eqref{chap2:eq:magnetic_tensor}, the equation for the total energy tensor $\rho V_i V_j + P_{ij} + B_i B_j$ is obtained as follows: 
\begin{align}
    &\partial_t (\rho V_i V_j + P_{ij} + B_i B_j) + \partial_\alpha (\rho V_i V_j V_\alpha + V_i P_{j \alpha} + V_j P_{\alpha i} + V_\alpha P_{ij} + \epsilon_{i \alpha \beta} B_j E_\beta + \epsilon_{j \alpha \beta} B_i E_\beta)  \nonumber\\
    &= B_j V_\alpha \partial_\alpha B_i + B_i V_\alpha \partial_\alpha B_j - B_\alpha V_i \partial_j B_\alpha - B_\alpha V_j \partial_i B_\alpha. 
    \label{chap2:eq:total_energy_tensor}
\end{align}

As explained above, our model does not treat the rotation of the pressure tensor directly to reduce the computational cost. To mimic isotropization/gyrotropization, the term $-\nu_{\rm{g}} (\bm{P} - \bm{P}_{\rm{g}})$ is added to the RHS of Eq.\eqref{chap2:eq:total_energy_tensor}. When the plasma approaches a gyrotropic state, $P_{\rm{g}} = P_\perp \bm{I} + (P_\parallel - P_\perp) \hat{\bm{b}} \hat{\bm{b}}$ where $\hat{\bm{b}} := \bm{B} / |\bm{B}|$, and $P_\parallel, P_\perp$ are the parallel and perpendicular pressures for the magnetic field. When the plasma approaches an isotropic state, $P_{\rm{g}} = P \bm{I}$ where $P = (P_\parallel + 2 P_\perp) / 3$. $\nu_{\rm{g}}$ is a free parameter representing the relaxation rate toward the isotropic/gyrotropic equilibrium state $\bm{P}_{\rm{g}}$.  

Combining the above results, the final set of 10-moment MHD equations is given as follows: 
\begin{align}
    &\partial_t \rho + \partial_\alpha (\rho V_\alpha) = 0, \\ 
    &\partial_t (\rho V_i) + \partial_\alpha \left( \rho V_i V_\alpha + P_{i \alpha} + \frac{B^2}{2} \delta_{i \alpha} - B_i B_\alpha \right) = 0, \\
    &\partial_t B_i + \partial_\alpha (B_i V_\alpha - B_\alpha V_i) = 0, \\ 
    &\partial_t (\rho V_i V_j + P_{ij} + B_i B_j) + \partial_\alpha (\rho V_i V_j V_\alpha + V_i P_{j \alpha} + V_j P_{\alpha i} + V_\alpha P_{ij} + \epsilon_{i \alpha \beta} B_j E_\beta + \epsilon_{j \alpha \beta} B_i E_\beta)  \nonumber\\
    &= B_j V_\alpha \partial_\alpha B_i + B_i V_\alpha \partial_\alpha B_j - B_\alpha V_i \partial_j B_\alpha - B_\alpha V_j \partial_i B_\alpha - \nu_{\rm{g}} (P_{ij} - P_{{\rm{g}}, ij}).  
\end{align}

In \citet{hirabayashi2016}, the conservative terms ($\partial_\alpha (\cdot)$ terms on the LHS) are solved using an HLL approximate Riemann solver for the 10-moment equation, while the non-conservative terms are calculated using a finite difference scheme with a limiter function. The isotropization/gyrotropization effect is treated analytically as $\bm{P}(t + \Delta_t) = \bm{P}_{\rm{g}} + (\bm{P}(t) - \bm{P}_{\rm{g}}) e^{-\nu_{\rm{g}} \Delta_t}$. To satisfy the isotropic/gyrotropic limits, the calculation for the total energy tensor is divided into the following steps: (1) advance the total energy tensor at one step without the $- \nu_{\rm{g}} (P_{ij} - P_{{\rm{g}}, ij})$ term; (2) calculate $P_{\rm{g}}$ using the updated pressure tensor; (3) advance the total energy tensor using the $- \nu_{\rm{g}} (P_{ij} - P_{{\rm{g}}, ij})$ term. 


\section{Overview of the finite difference scheme for conventional MHD with nonlinear filtering}\label{chap3:review_of_iijima_scheme}

We provide an overview of the finite-difference scheme with nonlinear filtering proposed by \citet{iijima2021}. The scheme evolves the internal energy equation instead of the total energy equation, thereby avoiding the occurrence of negative pressure caused by numerical errors, which can become particularly problematic in low-$\beta$ plasmas. Although the total energy equation is not solved directly, the scheme is designed to conserve the total energy within the accuracy of the temporal discretization by explicitly evaluating the numerical dissipation of kinetic and magnetic energies and transferring the dissipated energy into the internal energy. To conserve the momentum and accurately evaluate the numerical kinetic-energy dissipation, the convective term in the momentum equation is written in a skew-symmetric split form that ensures the momentum conservation and the appropriate kinetic-energy balance at the discrete level. The discretized governing equations are given by 
\begin{align}
    &\partial_t \rho + \delta_\alpha (\rho V_\alpha) = \delta_{1, \alpha} D_\alpha(\rho), \\ 
    &\partial_t (\rho V_i) + \frac{1}{2} \delta_\alpha (\rho V_i V_\alpha) + \frac{1}{2} \rho V_\alpha \delta_\alpha V_i + \frac{1}{2} V_i \delta_\alpha (\rho V_\alpha) + \delta_i p + B_\alpha \delta_i B_\alpha - B_\alpha \delta_\alpha B_i = \delta_{1, \alpha} D_\alpha(\rho V_i), \\
    &\partial_t B_i + \delta_\alpha (B_i V_\alpha - B_\alpha V_i) + \delta_i \psi = \delta_{1, \alpha} D_\alpha(B_i), \\
    &\partial_t e + \delta_\alpha ((e + p) V_\alpha) = V_\alpha \delta_\alpha p + \delta_{1, \alpha} D_\alpha(e) + Q_{{\rm{vis}}} + Q_{{\rm{res}}} + Q_{\psi}, \\
    &\partial_t \psi + c_\psi^2 \delta_\alpha B_\alpha = \frac{\psi}{\tau_\psi} + \delta_{1, \alpha} D_\alpha(\psi).
\end{align}
Here, $p$ is the scalar pressure and $e = p / (\gamma_{\rm{heat}} - 1)$ the internal energy where $\gamma_{\rm{heat}}$ is the heat capacity ratio. $D_\alpha(\cdot)$ denotes the artificial diffusive flux introduced through nonlinear filtering, and $Q_{\rm vis}$, $Q_{\rm res}$, and $Q_{\psi}$ denote the numerical viscous heating, numerical resistive heating, and heating associated with $\psi$, respectively. $\psi$ is the scalar variable introduced to control the $\nabla \cdot \bm{B}$ error, in what is known as the 9-wave method \cite{dedner2002}. $c_\psi$ is the propagation speed, and $\tau_\psi$ is the diffusion time scale of $\psi$. 

In this scheme, the the discrete spatial derivative 
$\delta_\alpha \Phi$ is defined as follows: 
\begin{align}
    &\delta_\alpha \Phi := \sum_{n = 1}^{N_{\rm{c}}} c_n \delta_{2n, \alpha} \Phi, \\
    &\delta_{n, \alpha} \Phi := \frac{\Phi(x_\alpha + n\Delta_\alpha / 2) - \Phi(x_\alpha - n\Delta_\alpha / 2)}{n\Delta_\alpha}, 
\end{align}
where $\Phi$ is some physical quantity, $\Delta_\alpha$ the spatial grid size in the $\alpha$ direction, and $n$ the stencil. The coefficients $c_n$ are determined as follows: $c_1 = 1$ for second-order accuracy; $c_1 = 4/3$ and $c_2 = -1/3$ for fourth-order accuracy; and $c_1 = 3/2$, $c_2 = -3/5$, and $c_3 = 1/10$ for sixth-order accuracy.

The diffusive flux is calculated using some limiter function as follows: 
\begin{align}
    (D_\alpha \Phi) \left( x + \frac{\Delta_\alpha}{2} \right) = \frac{c_{\rm{diff}} |a|}{2} \left( \Phi^{\rm{R}} \left( x + \frac{\Delta_\alpha}{2} \right) - \Phi^{\rm{L}} \left( x + \frac{\Delta_\alpha}{2} \right) \right), 
\end{align}
where $\Phi^{\rm{R}}$ and $\Phi^{\rm{L}}$ are the reconstructed values. In this paper, MUSCL with a minmod limiter \cite{vanleer1979} is used for the second-order finite difference scheme, and WENO5-JS \cite{jiang1996} is used for the fourth- and sixth-order finite difference schemes. $c_{\rm{diff}}$ is a free parameter, and its value is set to $\mathcal{O}(0.1 - 1)$. $a$ is the maximum characteristic speed in the system. 

Introducing the filtering flux produces numerical heating, which we add to the internal energy equation for energy consistency. For example, numerical heating from introducing $\delta_{1, \alpha} D_\alpha (\rho V_i)$ is calculated as follows: 
\begin{align}
    \partial_t \left( \frac{\rho V_\alpha V_\alpha}{2} \right)_{\rm{filter}} 
    &= V_\alpha (\partial_t (\rho V_\alpha))_{\rm{filter}} - \frac{V_\alpha V_\alpha}{2} (\partial_t \rho)_{\rm{filter}} \nonumber\\
    &= V_\alpha \delta_{1, \beta} D_\beta (\rho V_\alpha) - \frac{V_\alpha V_\alpha}{2} \delta_{1, \beta} D_\beta (\rho) \nonumber\\
    &= \delta_{1, \beta} \left( \overline{V_\alpha}^{1, \beta} D_\beta (\rho V_\alpha) - \overline{\frac{V_\alpha V_\alpha}{2}}^{1, \beta} D_\beta (\rho) \right) - \overline{D_\beta (\rho V_\alpha) \delta_{1, \beta} V_\alpha}^{1, \beta} + \overline{\frac{D_\beta (\rho)}{2} \delta_{1, \beta} (V_\alpha V_\alpha)}^{1, \beta}, 
\end{align}
where $(\cdot)_{\rm{filter}}$ denotes the term of the nonlinear filtering flux. Here, we define $\overline{\Phi}^{n, i}$ as follows:
\begin{align}
    \overline{\Phi}^{n, i} := \frac{\Phi(x_i + n \Delta_i / 2) + \Phi(x_i - n \Delta_i / 2)}{2}. 
\end{align}
In this derivation, we use the discretized Leibniz rule \cite{morinishi1998} as follows: 
\begin{align}
    \overline{\Psi \delta_{n, i} \Phi}^{n, i} + \Phi \delta_{n, i} \Psi = \delta_{n, i} (\overline{\Phi}^{n, i} \Psi).
\end{align} 
The last two terms (denoted $-Q_{\rm{vis}}$) should be subtracted from the energy equation to avoid an inconsistency in the total energy. Applying a similar procedure to the magnetic energy $B_\alpha B_\alpha / 2$, we obtain 
\begin{align}
    &Q_{\rm{res}} = \overline{D_\beta (B_\alpha) \delta_{1, \beta} B_\alpha}^{1, \beta}, \\ 
    &Q_\psi = -\psi \delta_\alpha B_\alpha. 
\end{align}
We note that the discrete derivative for $Q_\psi$ is $\delta_{\alpha}$.


\section{Extension of the energy-consistent finite-difference scheme to 10-moment MHD}\label{chap4:extension_to_10-moment_MHD}

To extend the energy-consistent finite-difference scheme with nonlinear filtering described in Section \ref{chap3:review_of_iijima_scheme} to 10-moment MHD, we first derive the evolution equation for $P_{ij}$ from the 10-moment equations as follows:
\begin{align}
    \partial_t P_{ij} + \partial_\alpha (V_i P_{j \alpha} + V_j P_{\alpha i} + V_\alpha P_{ij}) = V_i \partial_\alpha P_{j \alpha} + V_j \partial_\alpha P_{i \alpha}. 
\end{align}
By replacing the spatial derivatives with the discrete spatial derivative operators $\delta_\alpha$ and adding nonlinear filtering fluxes to the right-hand side of each equation, we obtain the discretized 10-moment MHD equations as follows:
\begin{align}
    &\partial_t \rho + \delta_\alpha (\rho V_\alpha) = \delta_{1, \alpha} D_\alpha(\rho), \\ 
    &\partial_t (\rho V_i) + \frac{1}{2} \delta_\alpha (\rho V_i V_\alpha) + \frac{1}{2} \rho V_\alpha \delta_\alpha V_i + \frac{1}{2} V_i \delta_\alpha (\rho V_\alpha) + \delta_\alpha P_{i \alpha} + B_\alpha \delta_i B_\alpha - B_\alpha \delta_\alpha B_i = \delta_{1, \alpha} D_\alpha(\rho V_i), \\
    &\partial_t B_i + \delta_\alpha (B_i V_\alpha - B_\alpha V_i) + \delta_i \psi = \delta_{1, \alpha} D_\alpha(B_i), \\
   &\partial_t P_{ij} + \delta_\alpha (V_i P_{j \alpha} + V_j P_{\alpha i} + V_\alpha P_{ij}) \nonumber\\
    &= V_i \delta_\alpha P_{j \alpha} + V_j \delta_\alpha P_{i \alpha} + \delta_{1, \alpha} D_\alpha(P_{ij}) + 2Q_{{\rm{vis}}, ij} + 2Q_{{\rm{res}}, ij} + 2Q_{\psi, ij} - \nu_{\rm{g}} (P_{ij} - P_{{\rm{g}}, ij}), \\
    &\partial_t \psi + c_\psi^2 \delta_\alpha B_\alpha = \frac{\psi}{\tau_\psi} + \delta_{1, \alpha} D_\alpha(\psi).
\end{align} 

Taking the trace of Eq.~\eqref{chap2:eq:total_energy_tensor}, we find that $\rho V_\alpha V_\alpha + P_{\alpha \alpha} + B_\alpha B_\alpha$ should be conserved. In the isotropic limit, this quantity corresponds to twice the conventional total energy density. To satisfy this conservation, the heating terms are specified as follows: 
\begin{align}
    &Q_{{\rm vis}, \alpha \alpha} = \overline{D_\beta (\rho V_\alpha) \delta_{1, \beta} V_\alpha}^{1, \beta} - \overline{\frac{D_\beta (\rho)}{2} \delta_{1, \beta} (V_\alpha V_\alpha)}^{1, \beta}, \\
    &Q_{\rm{res}, \alpha \alpha} = \overline{D_\beta (B_\alpha) \delta_{1, \beta} B_\alpha}^{1, \beta}, \\ 
    &Q_{\psi, \alpha \alpha} = -\psi \delta_\alpha B_\alpha. 
\end{align}
The partition of the numerical heating among the components of the pressure tensor can be explicitly prescribed. This flexibility represents a distinct advantage over schemes based on approximate Riemann solvers, in which the partition of numerical heating is generally determined implicitly by the numerical scheme and cannot be directly controlled. In our fiducial setup, we assume equipartition of the heating among the three diagonal components,
\begin{align}
    Q_{{\rm vis},ii} = Q_{{\rm vis},jj} = Q_{{\rm vis},kk} = \frac{1}{3}Q_{{\rm vis},\alpha\alpha},
\end{align}
while the off-diagonal components are set to zero. $Q_{{\rm res}, ij}$ and $Q_{\psi, ij}$ are calculated in the same way.


\section{Test problems}\label{chap5:test_problems}

In this section, we show some test simulation results. At first, we check the convergence and energy conservation properties by performing the circularly polarized Alfv\'en wave propagation. Next, we compare the test simulation results with previous studies \cite{hirabayashi2016, iijima2021, luo2023, bhoriya2024, singh2025}. Finally, we focus on the temperature anisotropy effects in high beta systems (CGL closure becomes numerically unstable), and the numerical stability for relatively low beta systems (approximate Riemann solver easily produces negative pressure to crash numerical calculations). 

In all test simulations, we employ the third-order strong-stability-preserving Runge--Kutta (SSP-RK3) method \citep{shu1988} for time integration. The isotropic and gyrotropic limits are numerically realized by setting the relaxation rate to a sufficiently large value, $\nu_{\rm g}=10^{20}$. In the isotropic limit, the target pressure tensor is defined as $\bm{P}_{\rm g}=P\bm{I}$, where $P=(P_\parallel+2P_\perp)/3$. In the gyrotropic limit, the target pressure tensor is given by $\bm{P}_{\rm g} =P_\perp\bm{I} +(P_\parallel-P_\perp)\hat{\bm{b}}\hat{\bm{b}}$, where $\hat{\bm{b}}:=\bm{B}/|\bm{B}|$. For simulations without isotropization/gyrotropization, we set $\nu_{\rm g}=0$. The divergence-cleaning timescale is set to $\tau_\psi=0.18/c_\mathrm{h}$, where $c_\mathrm{h}:={\rm CFL}\,\min(\Delta_i,\Delta_j,\Delta_k)/\Delta_t$. Here, $\mathrm{CFL}$ is the Courant–Friedrichs–Lewy number. 


\subsection{Circularly polarized Alfv\'en wave propagation in the gyrotropic limit}\label{chap5:circularly_poralized_alfven_wave}

The first test simulation is a circularly polarized Alfv\'en wave propagation to verify the numerical convergence of the proposed scheme. The initial conditions are given as follows: 
\begin{align*}
    &\rho = \rho_0, \\
    &\bm{V} = (0, \delta V_{\rm{A0}} \cos(2 \pi x), \delta V_{\rm{A0}} \sin(2 \pi x)), \\
    &\bm{B} = (B_0, \delta B_0 \cos(2 \pi x), \delta B_0 \sin(2 \pi x)), \\
    &P_{ij} = P_{\perp 0} \delta_{ij} + (P_{\parallel 0} - P_{\perp 0}) \times B_i B_j / |\bm{B}|^2. 
\end{align*}
$\delta$ is the circularly polarized Alfv\'en wave amplitude and set it to $\delta = 0.001$ for investigating the L1 convergence and $\delta = 0.1$ the energy conservation. $V_{\rm{A0}}$ is the Alfv\'en-wave eigenmode in an anisotropic plasma, which is given by $V_{\rm A0,anisotropic}=\sqrt{\epsilon}\,V_{\rm A0,isotropic}$, where $\epsilon := 1 - (P_\parallel - P_\perp) / |\bm{B}|^2$ and $V_{\rm A0,isotropic} := B_0 / \sqrt{\rho_0}$. We set $\rho_0 = 1, B_0 = 1$, and $P_{\perp 0} = 0.25$. Three cases with different $P_\parallel / P_\perp$ ratios ($0.5, 1.0$, and $2.0$) are investigated. The simulation box size is $1$, corresponding to a single wavelength. In calculating the nonlinear filtering flux, MUSCL reconstruction is used for the second-order scheme, and WENO5-JS reconstruction is used for fourth- and sixth-order schemes, respectively. Periodic boundary conditions are imposed in the $x$ direction.

\begin{figure}
	\centering
	\includegraphics[width=.8\textwidth]{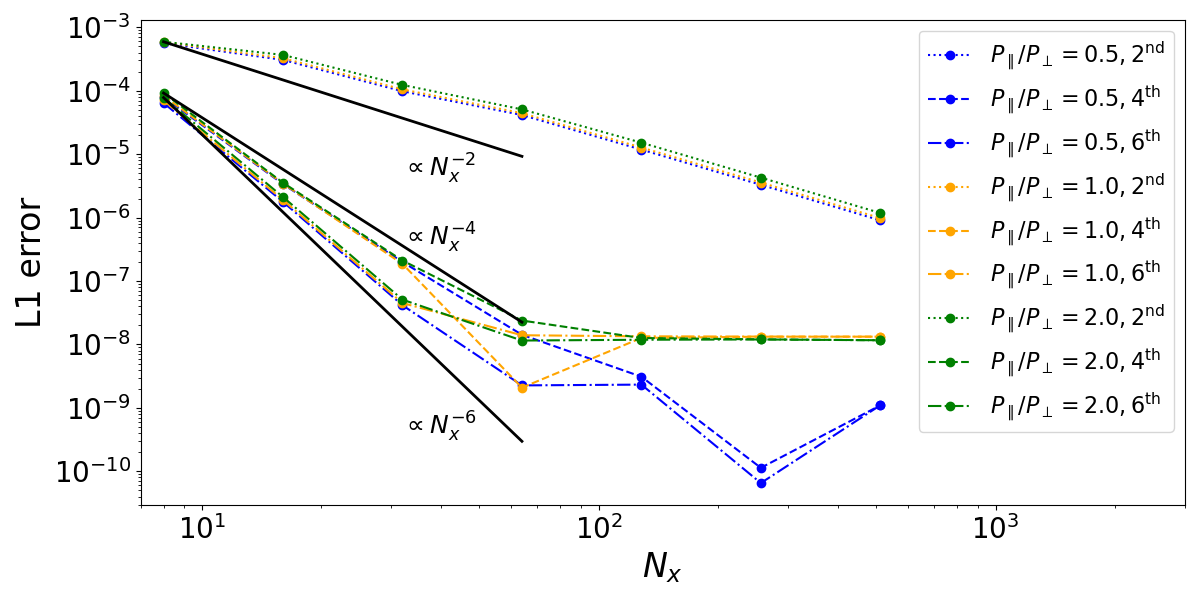}
	\caption{Convergence of the L1 error for $B_z$ for $P_\parallel / P_\perp = 0.5, 1.0, 2.0$ cases with second-, fourth-, and sixth-order finite differences. The solid lines show the second-, fourth-, and sixth-order references. The L1 error of the second-order scheme follows the expected scaling, while those of the fourth- and sixth-order schemes saturate around $N_x \gtrsim 64$.}
	\label{chap5:fig:circularly_polarized_alfven_wave:l1_error}
\end{figure}

Figure \ref{chap5:fig:circularly_polarized_alfven_wave:l1_error} shows the L1 error $\sum_{Nx} |B_z - B_{z, \rm{exact}}| / N_x$ after 1 cycle of propagating the circularly polarized Alfv\'en wave for different $P_\parallel / P_\perp$ value in the gyrotropic limit. The number of grid is changed from $N_x = 8$ to $N_x = 512$. The L1 convergence is satisfied when the second-order scheme with MUSCL is used. On the other hand, we find the saturation of convergence for fourth- and sixth-order schemes when the number of grid points becomes large. The L1 error saturates at the same level regardless of whether fourth- or sixth-order scheme is used. We also performed the same simulations using single (32-bit) precision and found that the behavior of the L1 error convergence is the same, indicating that the round-off error does not affect the result. Moreover, We performed the same simulations using a small wave amplitude of $\delta = 0.00001$ and found that the behavior is the same, indicating that the nonlinear evolution of the wave structure does not affect the result. We further performed the same simulations using a smaller CFL number and found that the behavior is the same. This indicates that the $\Delta_t$ does not affect the result, suggesting that the isotropization model is not the reason for the poor convergence. These results suggest that the saturation of the L1 error may originate from the numerical scheme proposed in this paper, such as the assumption of the equipartition of the numerical heating among the three diagonal components of the pressure tensor. 

\begin{figure}
	\centering
	\includegraphics[width=.8\textwidth]{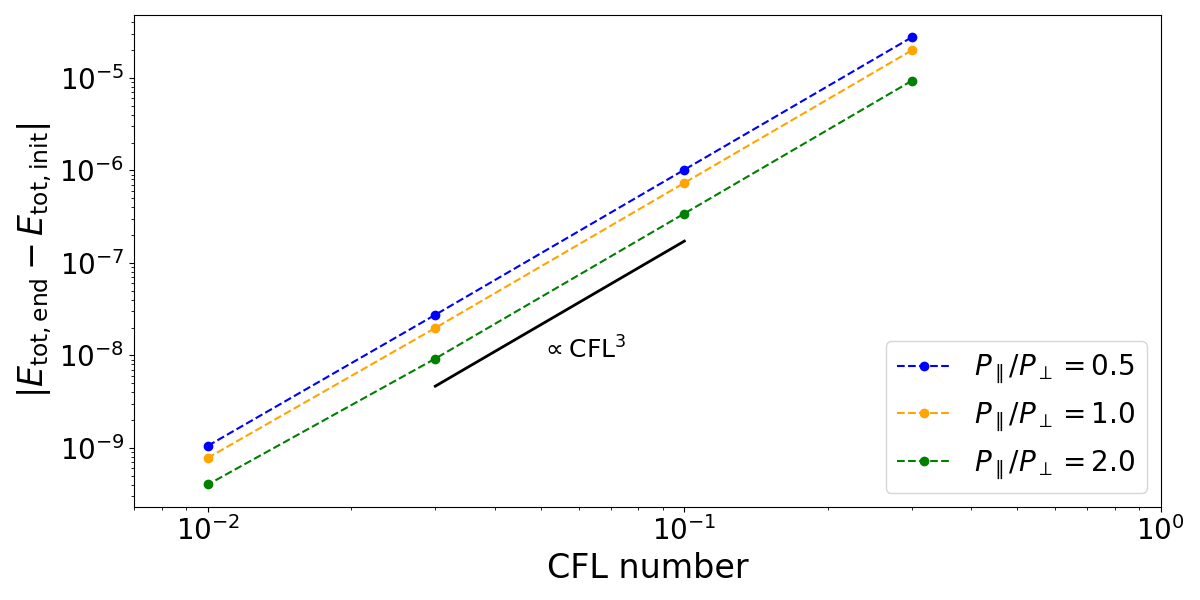}
	\caption{The error of the total energy conservation for $P_\parallel / P_\perp = 0.5, 1.0, 2.0$ cases with second-order finite difference. The solid line shows the third-order reference. The energy conservation errors follow the expected scalings in all cases.}
	\label{chap5:fig:circularly_polarized_alfven_wave:energy_conservation}
\end{figure}

Figure \ref{chap5:fig:circularly_polarized_alfven_wave:energy_conservation} shows the total energy error $|E_{\rm{tot,end}} - E_{\rm{tot,init}}|$ after 1 cycle of propagating the circularly polarized Alfv\'en wave for different $P_\parallel / P_\perp$ value in the gyrotropic limit with second-order scheme. Here, $E_{\rm{tot}}$ is defined as $(\rho V_\alpha V_\alpha + P_{\alpha \alpha} + B_\alpha B_\alpha) / 2$. The CFL number is changed from $0.01$ to $0.3$. The result shows that the energy conservation error scales as $\rm{CFL}^3$. We implement SSP-RK3 and it is consistent with the result. Since the trace of the pressure tensor remains unchanged when adding $\nu_{\rm{g}} (P_{ij} - P_{{\rm{g}}, ij})$, the isotropization model by construction does not affect the total energy conservation error. This is consistent with the fact that the energy conservation error scales with the order of accuracy of the time integrator (SSP-RK3 is used in this paper).


\subsection{Parametric decay instability}\label{chap5:parametric_decay}

The second test problem is the parametric decay instability (PDI), in which a finite-amplitude Alfv\'en wave becomes unstable and decays into a backward-propagating Alfv\'en wave and a compressive mode. Owing to the different linear growth rates predicted for the isotropic and gyrotropic limits, this problem enables independent validation of the two models through comparison of the numerical growth rates with the corresponding linear theories. The initial conditions are given as follows:
\begin{align*}
    &\rho = \rho_0, \\
    &\bm{V} = (0, \delta V_{\rm{A0}} \cos(2 N \pi x), \delta V_{\rm{A0}} \sin(2 N \pi x)), \\
    &\bm{B} = (B_0, \delta B_0 \cos(2 N \pi x), \delta B_0 \sin(2 N \pi x)), \\
    &P_{ij} = 0.5 \beta \delta_{ij}, 
\end{align*}
where the amplitude of the circularly polarized Alfv\'en wave is set to $\delta = 0.1$. Plasma beta is set to $\beta = 0.01$. Simulation box size is $1$ and resolved by $N_x = 1000$. $N$ is the number of wavelengths in the simulation domain and is set to $5$. We set $\rho_0 = 1$ and  $B_0 = 1$. $V_{\rm{A0}}$ is defined as $B_0 / \sqrt{\rho_0}$. We employ the sixth-order finite-difference scheme, with the nonlinear filtering flux computed using WENO5-JS reconstruction. Periodic boundary conditions are imposed in the $x$ direction. White noise with an amplitude of $0.001$ ($1\%$ of the amplitude of the Alfv\'en wave) is added to $\rho$ and $\bm{V}$ to trigger PDI. 

\subsubsection{Isotropic limit}

The dispersion relation for a circularly polarized Alfv\'en wave in conventional MHD has been derived in previous studies \cite{goldstein1978, derby1978} and is given by 
\begin{align}
    (\omega^2 - \bar{\beta} k^2) (\omega - k) ((\omega - k)^2 - 4) = \delta^2 k^2 (\omega^3 + \omega^2 k - 3 \omega + k). 
\end{align} 
Here, $\bar{\beta}$ is defined as $C_{\rm{S}}^2 / V_{\rm{A}}^2$ where $C_{\rm{S}}$ is the sound speed and $V_{\rm{A}}$ is the Alfv\'en speed. 

\begin{figure}
	\centering
	\includegraphics[width=.8\textwidth]{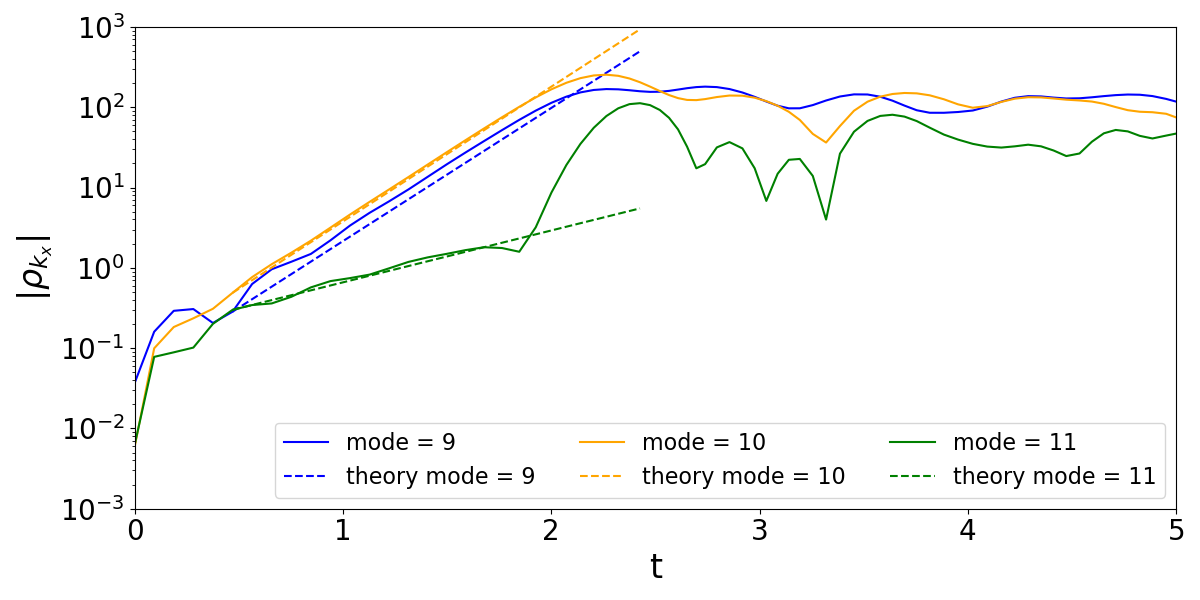}
	\caption{Comparison of the Fourier modes of $\rho$ in the isotropic limit (solid lines) with the linear growth rates predicted by the conventional MHD theory (dashed lines).}
	\label{chap5:fig:parametric_decay:linear_analysis_isotropic_limit}
\end{figure}

Figure \ref{chap5:fig:parametric_decay:linear_analysis_isotropic_limit} compares the time evolution of the Fourier amplitudes of the compressive mode for mode numbers $9$, $10$, and $11$ (solid lines), together with the corresponding linear growth predicted by conventional MHD theory (dashed lines). The simulation results of the Fourier mode 9, 10, and 11 agree well with the linear theory. This confirms that our model and scheme correctly reproduce conventional MHD theory in the isotropic limit. 

\subsubsection{Gyrotropic limit}

In the CGL-MHD model, the corresponding linear dispersion relation for a circularly polarized Alfv\'en wave has been derived in previous studies \cite{tenerani2017, saguchi2026} and is given by
\begin{align}
    &\left[\omega^2 - \tilde\beta k^2 \left(1 + \frac{\delta^2 \xi}{3}\right)\right]
    \left\{
        (\omega - k)\left[(\omega + k)^2 - 4\right]
        + \frac{\tilde\beta \delta^2 (\xi - 4)}{3(1 + \delta^2)}
        \left[(k^2 + 1)\omega + k(k^2 - 3)\right]
    \right\} \nonumber\\
    &= \epsilon^2 k^2 \left[1 - \frac{\tilde\beta(3 - \xi - \delta^2)}{3(1 + \delta^2)}\right]
    \left\{
        \omega^3 + \omega^2 k - 3\omega + k
        - \frac{\tilde\beta(3 - \xi)}{3}\left[(k^2 + 1)\omega + k(k^2 - 3)\right]
    \right\}, \nonumber\\
    &\tilde\beta = \frac{3\beta_\parallel}{2 \left( 1 + \delta^2 + \beta_\parallel (\xi - 1) / 2 \right)}, 
\end{align}
where $\xi := T_\perp / T_\parallel$ and $\beta_\parallel = P_\parallel / (B^2 / 2)$.

\begin{figure}
	\centering
	\includegraphics[width=.8\textwidth]{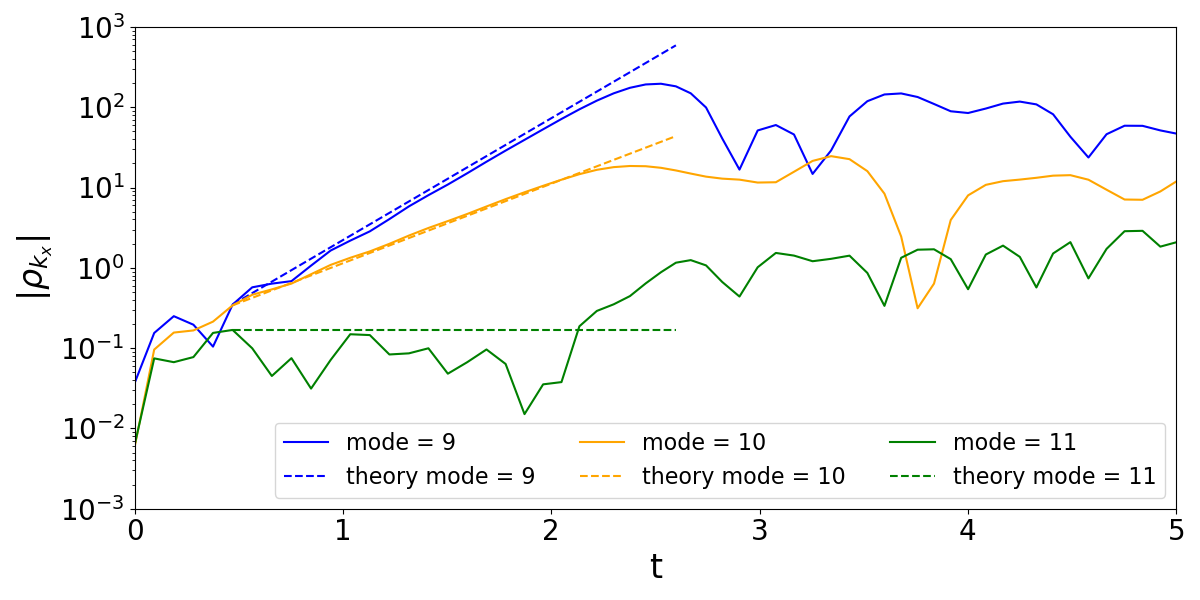}
	\caption{Comparison of the Fourier modes of $\rho$ in the gyrotropic limit (solid lines) with the linear growth rates predicted by the CGL-MHD theory (dashed lines).}
	\label{chap5:fig:parametric_decay:linear_analysis_gyrotropic_limit}
\end{figure}

Figure \ref{chap5:fig:parametric_decay:linear_analysis_gyrotropic_limit} shows the time evolution of the Fourier amplitudes of the compressive mode for mode numbers $9$, $10$, and $11$ (solid lines), together with the corresponding linear growth predicted by CGL-MHD theory (dashed lines). The simulation results agree well with the linear theory. In contrast to the isotropic limit, where the growth rates of modes $9$ and $10$ are nearly identical, the gyrotropic limit shows a clear difference between the growth rates of these two modes. Furthermore, modes with wavenumbers smaller than that of mode $10$ remain essentially unexcited, in agreement with the linear theory. These results confirm that our model and numerical scheme correctly reproduce the linear predictions of CGL-MHD theory in the gyrotropic limit.


\subsection{Shock tube problem}\label{chap5:shock_tube_problem}

The third test simulation is the shock tube problem proposed by \citet{brio1988}. This test simulation verifies the behavior around discontinuity or shock. High-order numerical schemes often generate grid-scale numerical oscillations near discontinuities. This test therefore provides a useful assessment of the ability of the present scheme to preserve monotonicity in the presence of strong discontinuities. The initial conditions are given as follows:
\begin{align*}
    &\rho_{\rm{L}} = 1, \ \rho_{\rm{R}} = 0.125, \\
    &\bm{V}_{\rm{L}} = (0, 0, 0), \ \bm{V}_{\rm{R}} = (0, 0, 0), \\ 
    &\bm{B}_{\rm{L}} = (0.75, 1, 0), \ \bm{B}_{\rm{R}} = (0.75, -1, 0), \\ 
    &P_{{\rm{L}}, ij} = \delta_{ij}, \ P_{{\rm{R}}, ij} = 0.1 \delta_{ij}. 
\end{align*}
Here, left half of the simulation box is denoted as $\rm{L}$ and right $\rm{R}$. Simulation box size is $1$ and it is resolved $256$ grids. In calculating the nonlinear filtering flux, MUSCL reconstruction is used for the second-order scheme, and WENO5-JS reconstruction is used for fourth- and sixth-order schemes, respectively. Free boundary conditions are imposed in the $x$ direction.

\subsubsection{Isotropic limit}\label{chap5:shock_tube_problem:isotropic_limit}

\begin{figure}
	\centering
	\includegraphics[width=\textwidth]{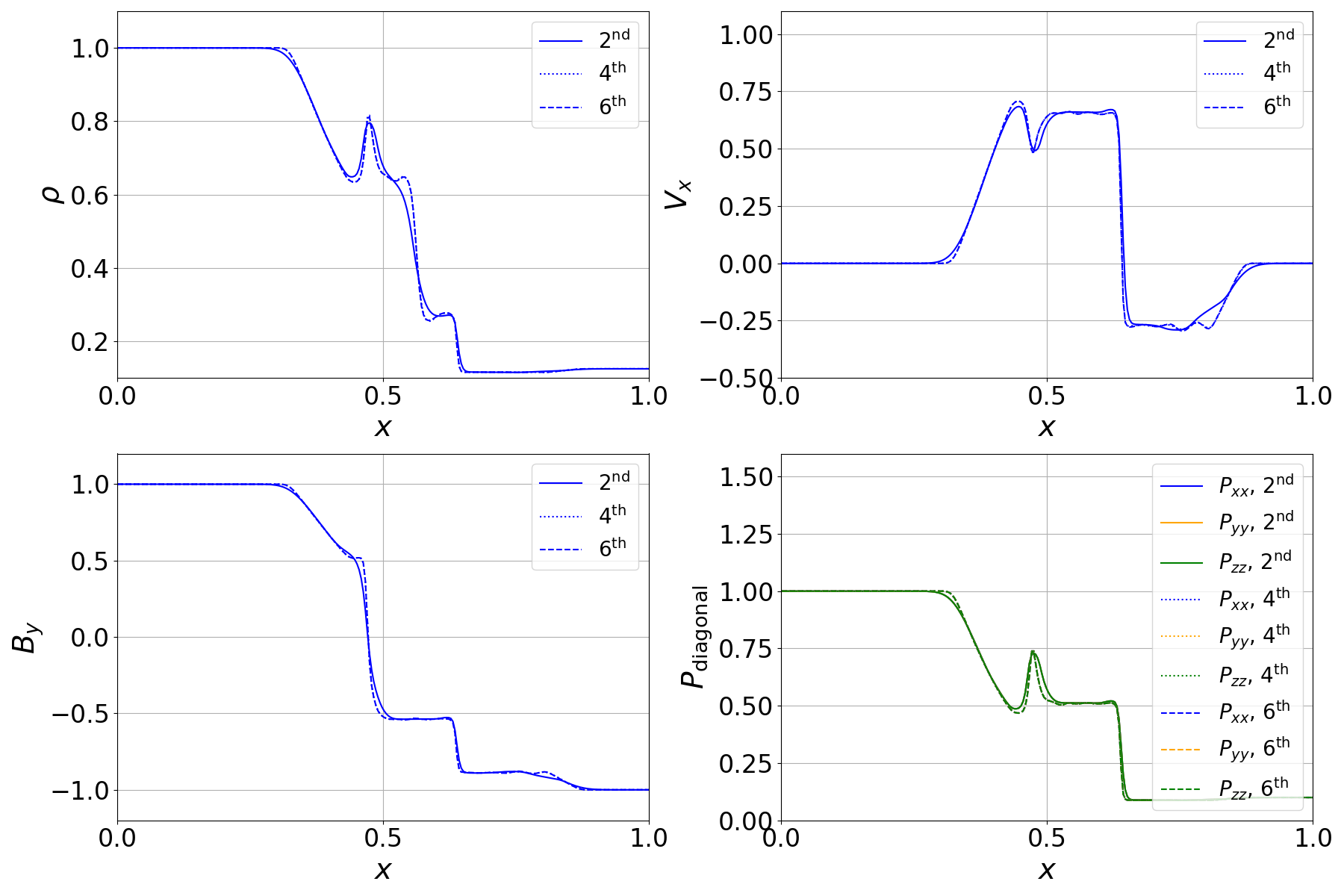}
	\caption{Snapshot of the \citet{brio1988} shock tube problem in the isotropic limit using second- (solid), fourth- (dotted), and sixth-order (dashed) finite differences at $t = 0.1$. $\rho$ (upper left), $V_x$ (upper right), $B_y$ (lower left), and $P_{\rm diagonal}$ (lower right) are shown. Here, blue corresponds to $P_{xx}$, orange to $P_{yy}$, and green to $P_{zz}$. The curves of $P_{xx}, P_{yy}$, and $P_{zz}$ overlap. Sharper structures appear near the discontinuities in the fourth- and sixth-order cases, while numerical oscillations appear around $x = 0.6$--$0.8$ in $V_x$.}
	\label{chap5:fig:shock_tube_problem:isotropic_limit}
\end{figure}

Figure \ref{chap5:fig:shock_tube_problem:isotropic_limit} shows snapshots of the shock tube problem in the isotropic limit using second-, fourth-, and sixth-order schemes at $t = 0.1$. A comparison among the different spatial orders shows that the contact discontinuity at $x \sim 0.6$ is more sharply resolved with the fourth- and sixth-order finite-difference schemes than with the second-order scheme. The resolution of the discontinuity is, however, nearly identical between the fourth- and sixth-order cases. This is because the numerical thickness of the discontinuity is primarily controlled by the nonlinear-filtering flux rather than by the order of the central finite difference. Consequently, the advantage of the sixth-order scheme over the fourth-order scheme appears mainly in smooth regions of the solution. This improved resolution comes at the cost of increased susceptibility to numerical oscillations. In the fourth- and sixth-order cases, oscillatory structures appear around the fast rarefaction wave at $x \sim 0.8$ and the slow shock at $x = 0.6$--$0.7$, whereas they are less pronounced in the second-order case. Previous work has shown that these oscillations can be reduced by increasing $c_{\rm diff}$ \cite{iijima2021}. The diagonal components of the pressure tensor, $P_{xx}$, $P_{yy}$, and $P_{zz}$, remain nearly identical, while the off-diagonal component satisfies $P_{xy} \sim 0$. These results demonstrate that the isotropization term operates as intended.

\subsubsection{Gyrotropic limit}\label{chap5:shock_tube_problem:gyrotropic_limit}

\begin{figure}
	\centering
	\includegraphics[width=\textwidth]{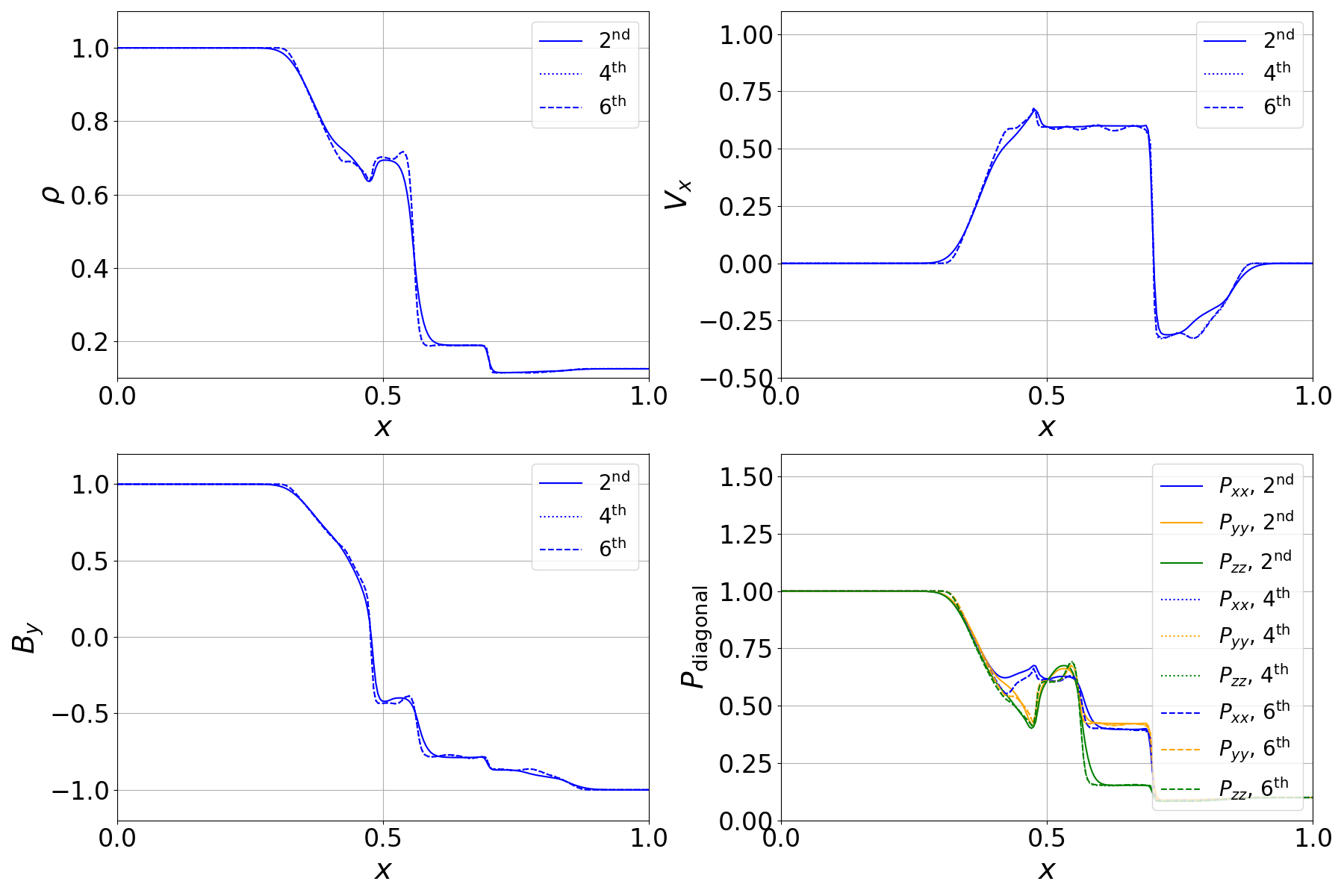}
	\caption{Snapshot of the \citet{brio1988} shock tube problem in the gyrotropic limit using second- (solid), fourth- (dotted), and sixth-order (dashed) finite differences at $t = 0.1$. $\rho$ (upper left), $V_x$ (upper right), $B_y$ (lower left), and $P_{\rm diagonal}$ (lower right) are shown. Here, blue corresponds to $P_{xx}$, orange to $P_{yy}$, and green to $P_{zz}$. The curves of $P_{xx}, P_{yy}$, and $P_{zz}$ differ. Sharper structures appear near the discontinuities in the fourth- and sixth-order cases, while numerical oscillations appear around $x = 0.6$--$0.8$ in $V_x$.}
	\label{chap5:fig:shock_tube_problem:gyrotropic_limit}
\end{figure}

Figure \ref{chap5:fig:shock_tube_problem:gyrotropic_limit} shows snapshots of the shock tube problem in the gyrotropic limit using second-, fourth-, and sixth-order schemes at $t = 0.1$. The pressure tensor exhibits the behavior expected in the gyrotropic limit: $P_{xx}$, $P_{yy}$, and $P_{zz}$ take different values. The fourth- and sixth-order schemes also resolve a sharp density discontinuity at $x \sim 0.4$, whereas this structure is substantially smoothed in the second-order result. As reported by \cite{hirabayashi2016}, the pressure profiles around the contact discontinuity ($x \sim 0.6$) and the slow shock ($x \sim 0.7$) differ from those in the isotropic limit and are qualitatively consistent with the expected CGL behavior. Since the magnetic field lies in the $x$--$y$ plane, $P_{zz}$ corresponds to the perpendicular pressure. It changes significantly across the contact discontinuity but only weakly across the slow shock, suggesting that the contact discontinuity primarily enhances $P_\perp$, whereas the slow shock primarily enhances $P_\parallel$ toward the center of the simulation domain ($x = 0.5$).

This interpretation is consistent with the CGL adiabatic invariants, $P_\perp/(\rho B) = \mathrm{const.}$ and $P_\parallel B^2/\rho^3 = \mathrm{const.}$ Across the contact discontinuity toward the center, $\rho$ increases by a factor of $3$ (from $0.2$ to $0.6$), while $B$ decreases by a factor of $\sqrt{2}$ because $B_y$ decreases from $-0.8$ to $-0.4$, with $B_x$ and $B_z$ remaining constant. The first adiabatic invariant therefore predicts an increase in $P_\perp$. Across the slow shock toward the center, $\rho$ increases and $B$ decreases, so the second adiabatic invariant predicts an increase in $P_\parallel$. Taken together, these results show that the proposed scheme reproduces the expected gyrotropic response and captures the CGL shock tube behavior at least qualitatively.

\subsubsection{Without isotropization/gyrotropization}\label{chap5:shock_tube_problem:without_isotropization_and_gyrotropization}

\begin{figure}
	\centering
	\includegraphics[width=\textwidth]{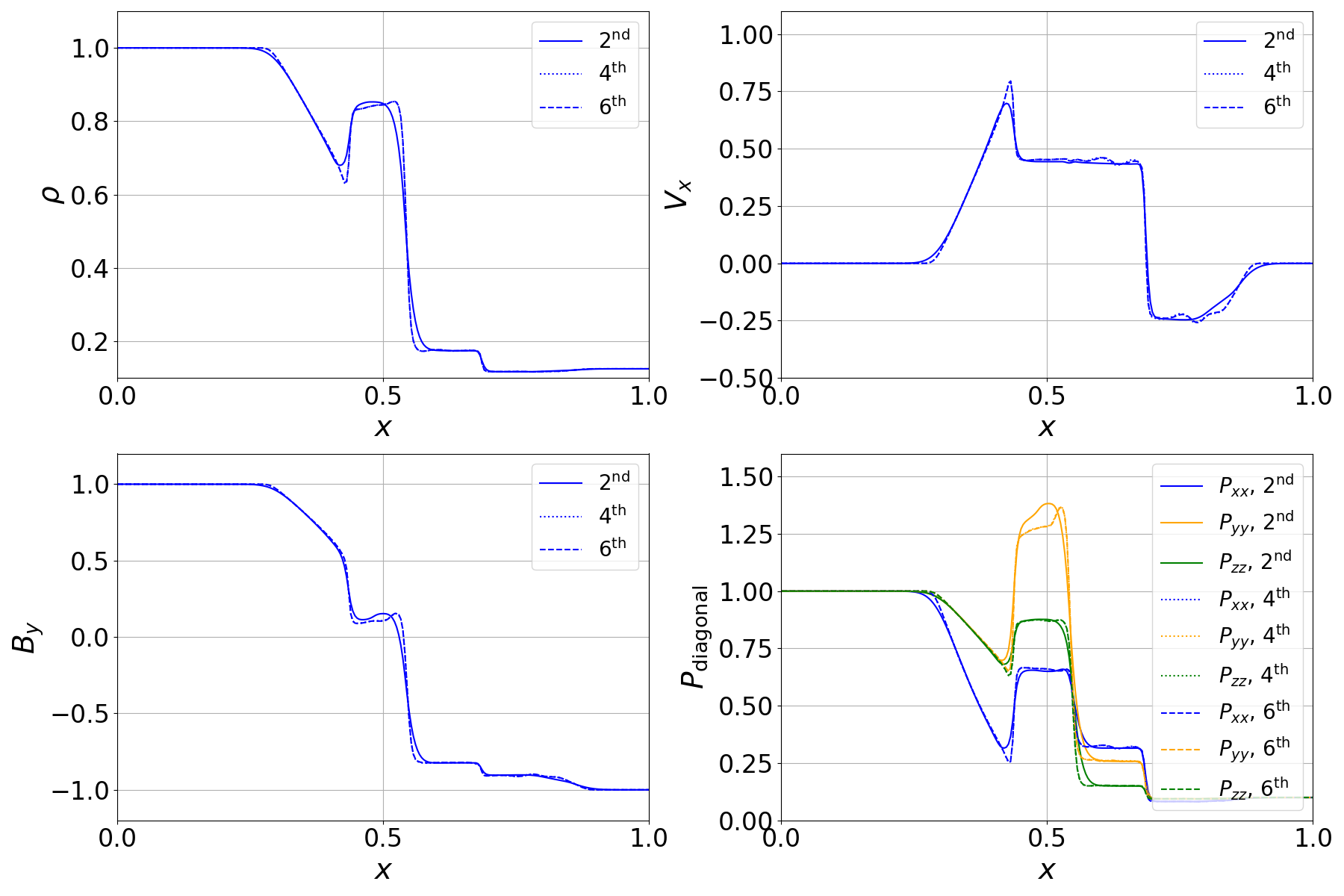}
	\caption{Snapshot of the \citet{brio1988} shock tube problem without isotropization/gyrotropization using second- (solid), fourth- (dotted), and sixth-order (dashed) finite differences at $t = 0.1$. $\rho$ (upper left), $V_x$ (upper right), $B_y$ (lower left), and $P_{\rm diagonal}$ (lower right) are shown. Here, blue corresponds to $P_{xx}$, orange to $P_{yy}$, and green to $P_{zz}$. Sharper structures appear near the discontinuities in the fourth- and sixth-order cases, while numerical oscillations appear around $x = 0.6$--$0.8$ in $V_x$.}
	\label{chap5:fig:shock_tube_problem:without_isotropization_and_gyrotropization}
\end{figure}

Figure \ref{chap5:fig:shock_tube_problem:without_isotropization_and_gyrotropization} shows snapshots of the shock tube problem without isotropization/gyrotropization using second-, fourth-, and sixth-order schemes at $t = 0.1$. In this case, all components of the pressure tensor evolve independently, without being constrained toward either an isotropic or a gyrotropic state. The solution contains rarefaction, compound-wave, contact discontinuity, and shock-like structures at approximately the same locations as those reported in the literature \cite{hirabayashi2016}. The profiles of the principal fluid variables and pressure tensor components also agree qualitatively with their results. Although no exact reference solution is available for the full 10-moment system, this agreement provides a useful consistency check between independent numerical implementations. 

As in the isotropic and gyrotropic cases, the fourth- and sixth-order schemes resolve sharper spatial structures than the second-order scheme, while also exhibiting more pronounced numerical oscillations near discontinuities. The fourth- and sixth-order results are nearly indistinguishable in the discontinuous regions, because their effective resolution is primarily controlled by the nonlinear-filtering flux. 


\subsection{Orszag-Tang vortex problem}\label{chap5:orszag-tang}

The fourth test simulation is the Orszag-Tang vortex problem proposed by \citet{orszag1979}. This problem is a standard multi-dimensional benchmark for assessing the robustness and spatial resolution of MHD schemes in flows involving interacting shocks and complex small-scale structures, as well as their ability to control numerical errors in $\nabla \cdot \bm{B}$. The initial conditions are given by 
\begin{align*}
    &\rho = \gamma_{\rm{heat}}^2, \\
    &\bm{V} = (-\sin(y), \sin(x), 0), \\ 
    &\bm{B} = (-\sin(y), \sin(2x), 0), \\ 
    &P_{ij} = \gamma_{\rm{heat}} \delta_{ij}, 
\end{align*}
where $\gamma_{\rm{heat}} = 5 / 3$. The simulation domain has dimensions of $2\pi \times 2\pi$ and is resolved with $N_x \times N_y = 256 \times 256$ grid points. In calculating the nonlinear filtering flux, MUSCL reconstruction is used for the second-order scheme, and WENO5-JS reconstruction is used for fourth- and sixth-order schemes, respectively. Periodic boundary conditions are imposed in the $x$ and $y$ directions.

\subsubsection{Isotropic limit}\label{chap5:orszag-tang:isotropic_limit}

\begin{figure}
	\centering
	\includegraphics[width=\textwidth]{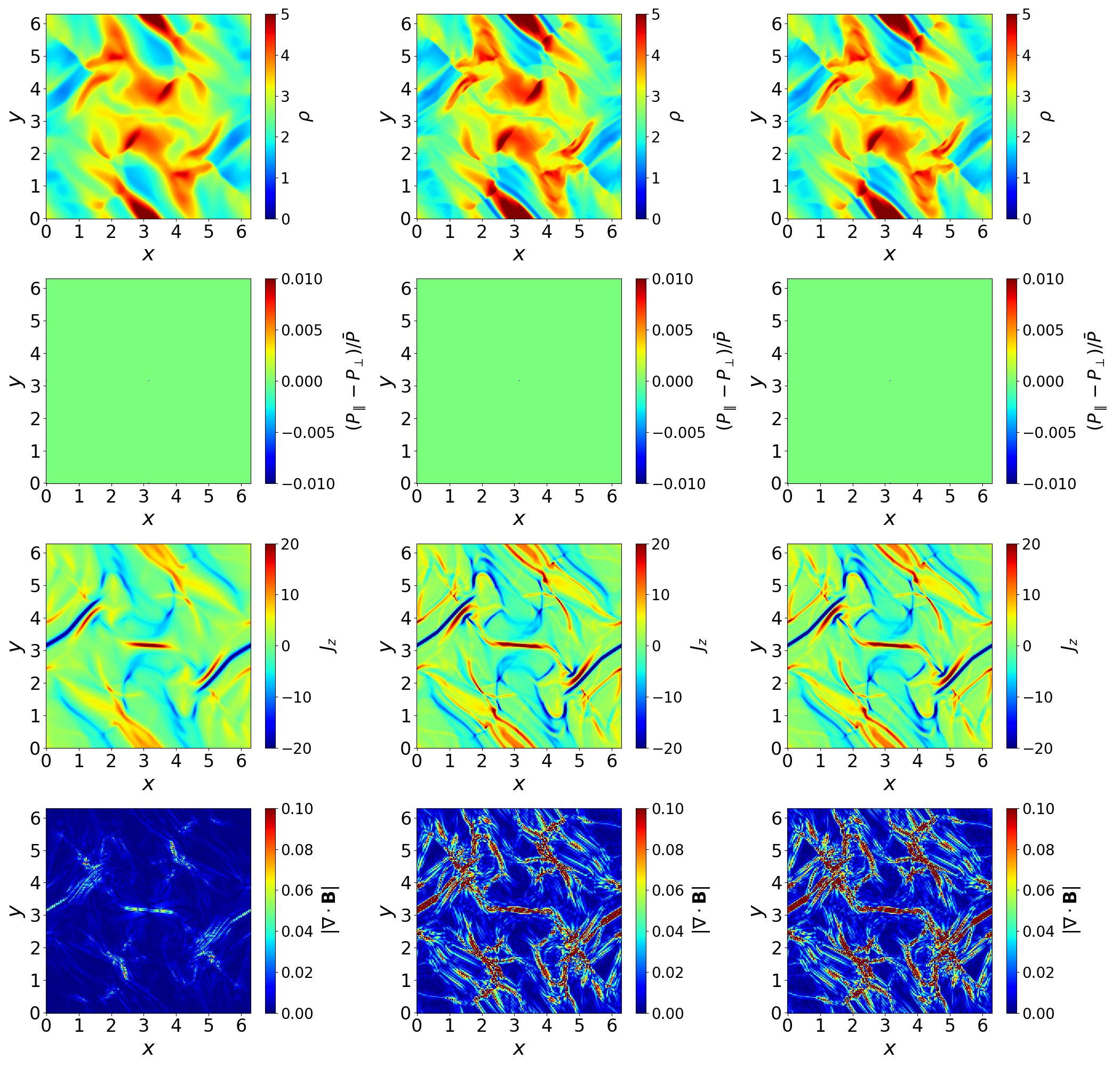}
	\caption{Snapshot of the \citet{orszag1979} vortex problem in the isotropic limit using second- (left), fourth- (center), and sixth-order (right) finite differences at $t = 4.0$. $\rho$, $(P_\parallel - P_\perp) / \bar{P}$, $J_z$ and $|\nabla \cdot \bm{B}|$ are shown. Here, $\bar{P}$ is defined as $(P_\parallel + 2 P_\perp) / 3$. $(P_\parallel - P_\perp) / \bar{P}$ is almost zero in all three cases. Small-scale structures are smoothed out in the second-order case due to numerical dissipation, while they appear in the fourth- and sixth-order cases.}
	\label{chap5:fig:orszag-tang:isotropic_limit}
\end{figure}

Figure \ref{chap5:fig:orszag-tang:isotropic_limit} shows snapshots of $\rho$, $(P_\parallel - P_\perp) / \bar{P}$, $J_z$ and $|\nabla \cdot \bm{B}|$ in the isotropic limit using second-, fourth-, and sixth-order schemes at $t = 4.0$. Compared with the second-order result, small-scale structures can be seen in the higher-order results, indicating that a higher-order scheme is needed to avoid numerical dissipation and to resolve the small-scale turbulent structures. On the other hand, the local value of $|\nabla \cdot \bm{B}|$ becomes larger. We can reduce the $|\nabla \cdot \bm{B}|$ error by adjusting the free parameters $C_\psi$, $\tau_\psi$, and $c_\mathrm{h}$ of the 9-wave scheme. A Sweet-Parker current sheet appears at the center of the simulation domain in all of the simulation results, which is consistent with previous studies \cite{stone2008, miyoshi2011, amano2015, matsumoto2019}.

\subsubsection{With finite isotropization}\label{chap5:orszag-tang:with_isotropization}

\begin{figure}
	\centering
	\includegraphics[width=\textwidth]{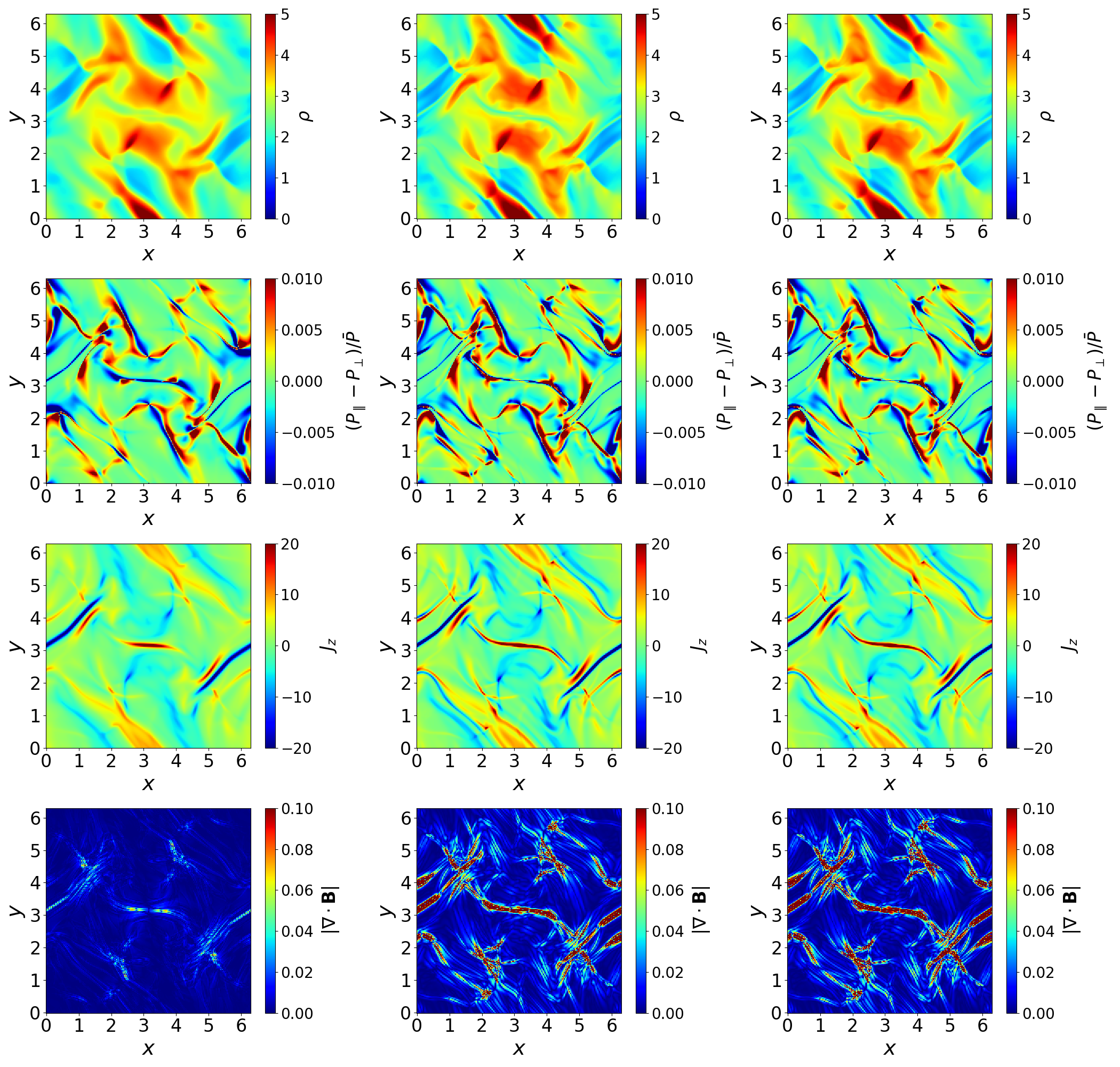}
	\caption{Snapshot of the \citet{orszag1979} vortex problem with finite isotropization effect using second- (left), fourth- (center), and sixth-order (right) finite differences at $t = 4.0$. $\rho$, $(P_\parallel - P_\perp) / \bar{P}$, $J_z$ and $|\nabla \cdot \bm{B}|$ are shown. Here, $\bar{P}$ is defined as $(P_\parallel + 2 P_\perp) / 3$. $(P_\parallel - P_\perp) / \bar{P}$ is nonzero in all three cases. Small-scale structures are smoothed out in the second-order case due to numerical dissipation, while they appear in the fourth- and sixth-order cases.}
	\label{chap5:fig:orszag-tang:with_isotropization}
\end{figure}

Figure \ref{chap5:fig:orszag-tang:with_isotropization} shows snapshots of $\rho$, $(P_\parallel - P_\perp) / \bar{P}$, $J_z$ and $|\nabla \cdot \bm{B}|$ with a finite isotropization rate of $\nu_{\rm{g}} = 100.0$, using second-, fourth-, and sixth-order schemes at $t = 4.0$. Here, $\bar{P}$ is defined as $(P_\parallel + 2 P_\perp) / 3$. As in the isotropic limit, small-scale structures are better resolved with the higher-order schemes. Compared with the isotropic-limit result, pressure anisotropy is clearly visible, especially around the discontinuities. The spatial profile of the pressure anisotropy agrees with that reported by \citet{bhoriya2024}. An elongated Sweet-Parker-like current sheet again forms near the center of the simulation box, as in the isotropic limit. The current sheet is compressed by the surrounding plasma, as indicated by the two mass density peaks above and below the current sheet (see the upper panels of Figure \ref{chap5:fig:orszag-tang:with_isotropization}), leading to an increase in $P_\perp$. This is consistent with the profile of $(P_\parallel - P_\perp) / \bar{P}$, which takes negative values within the current-sheet region. 


\subsection{Firehose instability in the gyrotropic limit}\label{chap5:firehose}

The fifth test simulation examines the firehose instability, for which the CGL-MHD system admits unstable modes. The criterion for the firehose instability is $\beta_\parallel - \beta_\perp > 2$ where $\beta_\parallel := 2 P_\parallel / |\bm{B}|^2$ and $\beta_\perp := 2 P_\perp / |\bm{B}|^2$. The initial conditions are given by 
\begin{align*}
    &\rho = \rho_0, \\
    &\bm{V} = (0, 0, 0), \\ 
    &\bm{B} = (B_0, 0, 0), \\ 
    &P_{xx} = P_{\parallel 0}, \ \ P_{yy} = P_{zz} = P_{\perp 0}, \ \ P_{\rm{off-diagonal}} = 0. 
\end{align*}
We set $\rho_0 = 1, B_0 = 1, P_{\parallel 0} = 4 \times 10^{10}$, and $P_{\perp 0} = 1 \times 10^{10}$. For these parameters, the initial state lies within the firehose-unstable regime, with $\beta_{\parallel 0} - \beta_{\perp 0} \gg 2$. The simulation domain is resolved with $N_x \times N_y = 100 \times 100$ grid points. We use a sixth-order scheme with a nonlinear filtering flux computed using WENO5-JS reconstruction. Periodic boundary conditions are imposed in the $x$ and $y$ directions. A circularly polarized Alfv\'en wave perturbation propagating in the $x$ direction, with an amplitude of $0.01 V_{\rm{A0}}$ for $\bm{V}$ and $0.01 B_0$ for $\bm{B}$ where $V_{\rm{A0}} := B_0 / \sqrt{\rho_0}$, is added to trigger the firehose instability. It consists of five wavelengths in the simulation domain. 

\begin{figure}
	\centering
	\includegraphics[width=\textwidth]{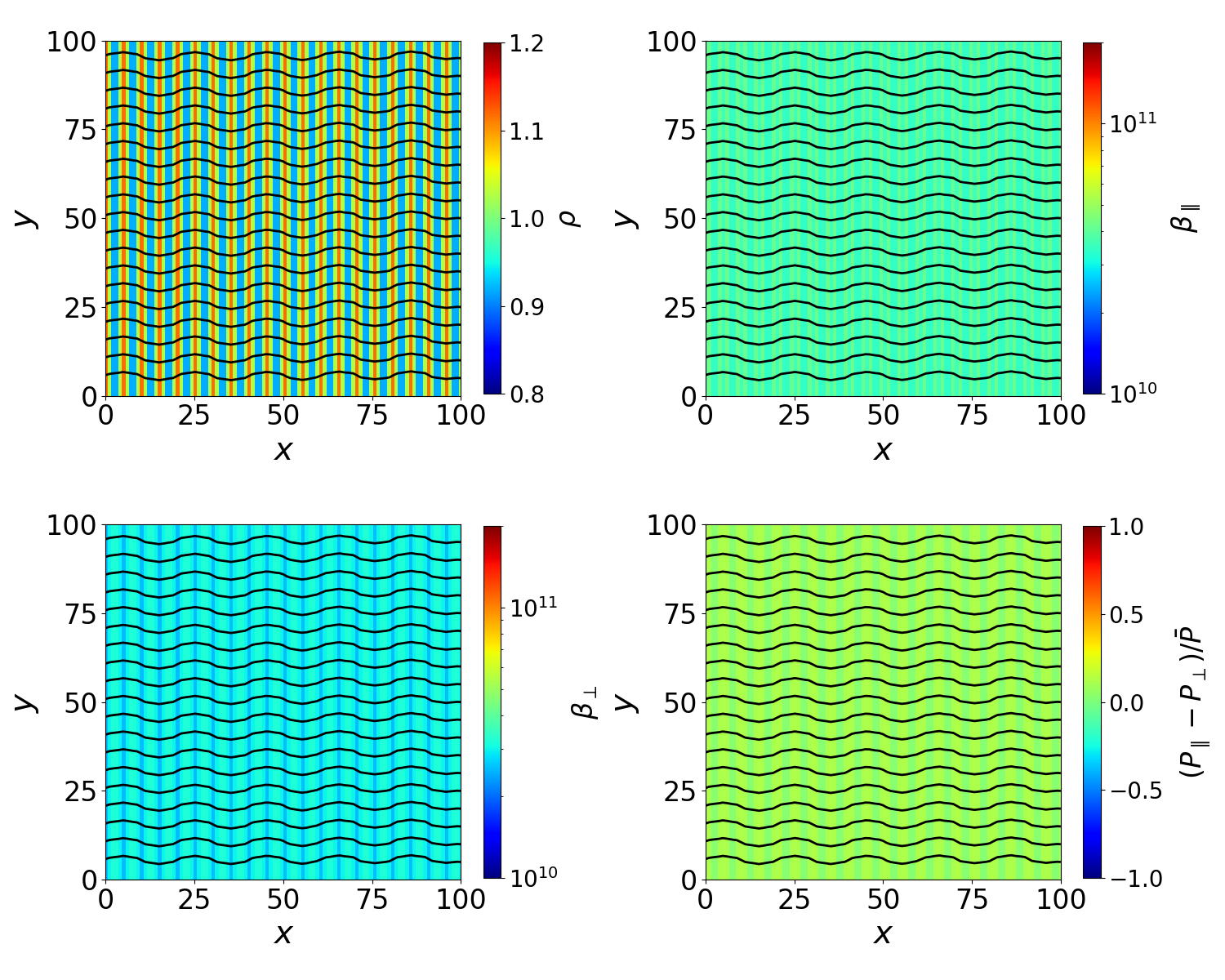}
	\caption{Snapshot of the firehose instability in the gyrotropic limit at $t = 0.0001$. $\rho$, $\beta_\parallel$, $\beta_\perp$ and $(P_\parallel - P_\perp) / \bar{P}$ are shown. Here, $\bar{P}$ is defined as $(P_\parallel + 2 P_\perp) / 3$. Black lines show the in-plane magnetic field. It can be seen that the magnetic field exhibits mode-5 structures, while shorter wavelength modes are also observed in these four snapshots.}
	\label{chap5:fig:firehose:gyrotropic_limit}
\end{figure}

Figure \ref{chap5:fig:firehose:gyrotropic_limit} shows snapshots of $\rho, \beta_\parallel, \beta_\perp$, and $(P_\parallel - P_\perp) / \bar{P}$ in the gyrotropic limit at $t = 0.0001$. Here, $\bar{P}$ is defined as $(P_\parallel + 2 P_\perp) / 3$. The firehose instability preferentially develops through modes with wavevectors parallel to the background magnetic field. The simulation develops stripe-like structures varying along the $x$ direction, accompanied by small-amplitude oscillations in $B_y$. The pressure anisotropy is also partially reduced as the instability develops. 

\begin{figure}
	\centering
	\includegraphics[width=.8\textwidth]{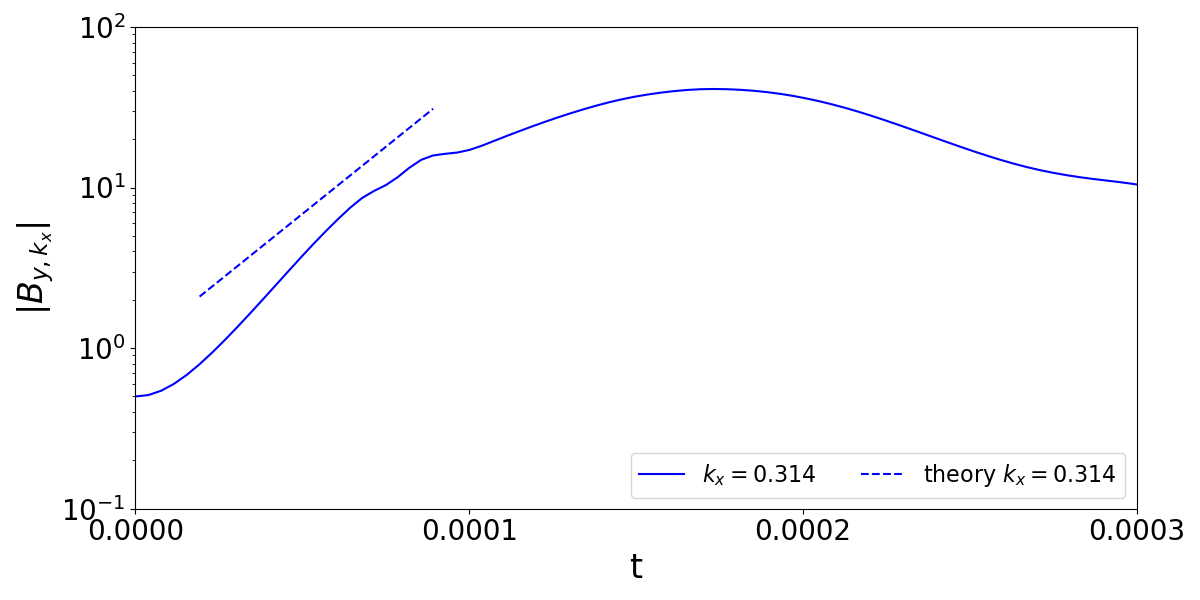}
	\caption{Comparison of the fastest-growing Fourier mode of $B_y$ (solid line) with the linear growth rate (dashed line). } 
	\label{chap5:fig:firehose:linear_analysis}
\end{figure}

Next, we compare the growth rate of the Fourier modes of $B_y$ with the linear theory. The growth rate of the firehose instability is given by 
\begin{align}
    \gamma = \frac{k_\parallel V_{\rm{A}}}{2} (\beta_\parallel - \beta_\perp - 2)^{1/2}. 
\end{align}
No constraint on the wavenumber appears in this equation, meaning that the fastest-growing mode is the grid-scale mode, since the largest $k_\parallel$ in numerical simulations is determined by the grid size. To avoid this problem, we introduce the five wavelength Alfv\'en wave perturbation in the initial conditions described above. The magnetic field lines shown in Figure \ref{chap5:fig:firehose:gyrotropic_limit} show mode-5 structures, consistent with this linear estimate. Figure \ref{chap5:fig:firehose:linear_analysis} shows the fastest-growing mode (solid line), together with the linear growth rate (dashed line). The wavenumber is $k_x \sim 3.14$, corresponding to a wavelength of $20$, consistent with the wavelength of the initial perturbation. The simulation results agree well with the linear theory, suggesting that the scheme proposed in this paper can correctly treat anisotropic effects in extremely high beta systems.


\subsection{Blast-wave problem}\label{chap5:blast_wave_problem}

The sixth test simulation is the blast-wave problem proposed by \citet{balsara1999}. This problem provides a stringent multidimensional test of numerical robustness because a large pressure contrast generates strong shocks that propagate through a strongly magnetized (low-beta) plasma. In particular, the extremely low ambient plasma beta adopted here allows us to examine whether the proposed scheme can maintain stable and positive pressure evolution. The initial conditions, following  \citet{iijima2021}, are given by 
\begin{align*}
    &\rho = \rho_0, \\
    &\bm{V} = (0, 0, 0), \\
    &\bm{B} = (B_0, 0, 0), \\
    &P_{ij} =
        \begin{cases}
        p_{\rm{inside}} \delta_{ij},    & \sqrt{x^2 + y^2} < 0.1, \\
        p_{\rm{outside}} \delta_{ij}, & \sqrt{x^2 + y^2} > 0.1. 
        \end{cases}
\end{align*}
We set $\rho_0 = 1$, $B_0 = 100/\sqrt{4\pi}$, $p_{\rm inside} = 10^3$, and $p_{\rm outside} = 10^{-7}$. The corresponding background plasma beta is $\mathcal{O}(10^{-10})$. The simulation domain has dimensions of $1.0 \times 1.0$ and is resolved with $N_x \times N_y = 200 \times 200$ grid points. In calculating the nonlinear filtering flux, MUSCL reconstruction and second-order scheme are used. Free boundary conditions are imposed in the $x$ and $y$ directions. 

\subsubsection{Isotropic limit}\label{chap5:blast_problem:isotropic_limit}

\begin{figure}
	\centering
	\includegraphics[width=\textwidth]{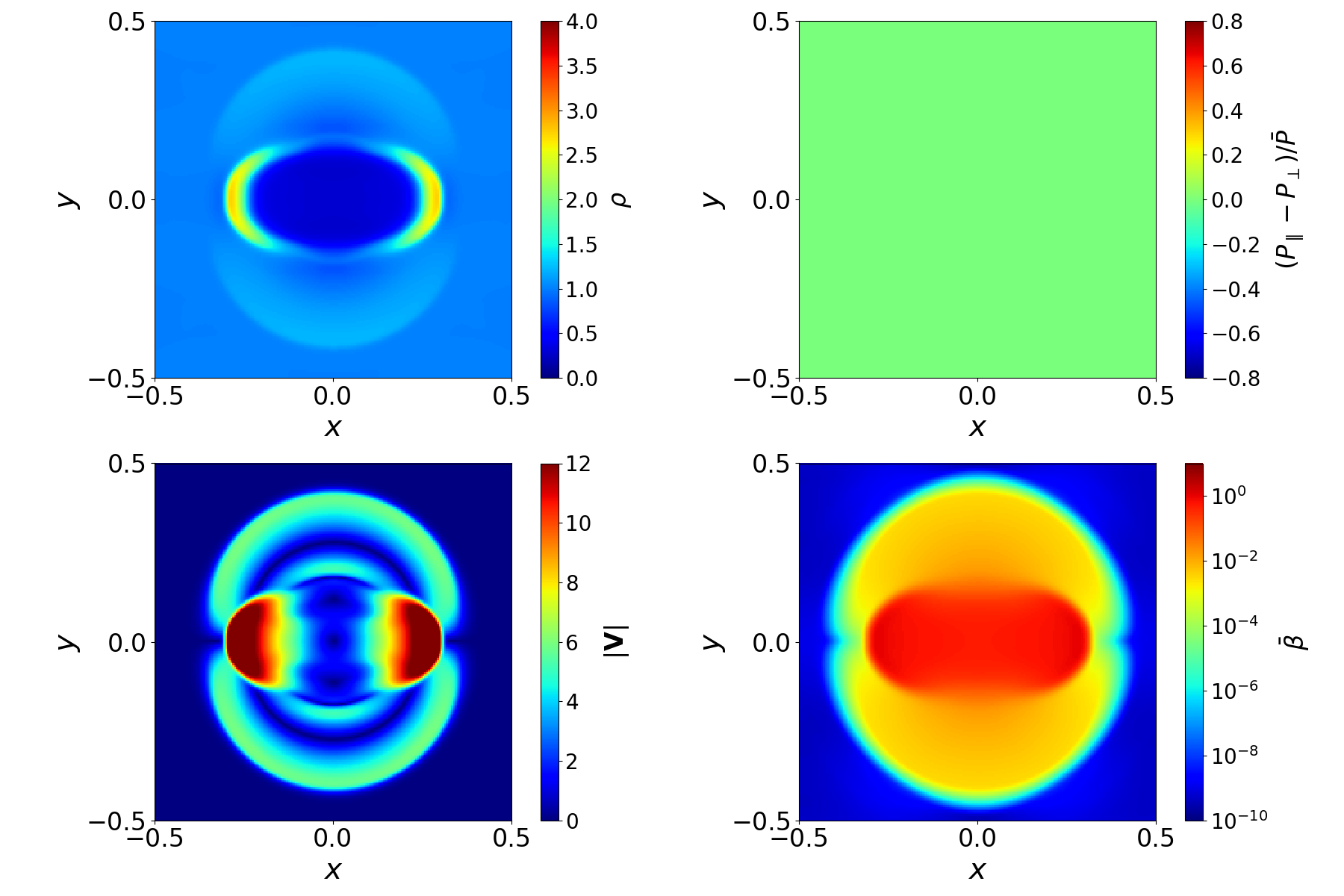}
	\caption{Snapshot of the blast problem in extremely low beta system in the isotropic limit at $t = 0.0090$. $\rho, (P_\parallel - P_\perp) / \bar{P}, |\bm{V}|, \bar{\beta}$ are shown. $\bar{\beta}$ is defined as $2 \bar{P} / |\bm{B}|^2$ where $\bar{P} := (P_\parallel + 2 P_\perp) / 3$. A pair of strong jets propagates along the $x$ direction, and the central core also elongates along the $x$ direction.}
	\label{chap5:fig:blast_problem:isotropic_list}
\end{figure}

Figure \ref{chap5:fig:blast_problem:isotropic_list} shows snapshots of $\rho$, $(P_\parallel - P_\perp) / \bar{P}$, $|\bm{V}|$, and $\bar{\beta}$ in the isotropic limit using the second-order scheme at $t = 0.0090$. Here, $\bar{P}$ is defined as $(P_\parallel + 2 P_\perp) / 3$ and $\bar{\beta}$ as $\bar{P} / (B^2 / 2)$. The fourth- and sixth-order schemes fail within a few time steps; therefore, only the second-order result is presented here. Although this reveals a limitation of the present higher-order implementation, the second-order scheme remains stable even in an extremely low beta system with $\beta \sim 10^{-10}$. This demonstrates the advantage of evolving the pressure tensor directly, since schemes that recover the thermal pressure from the total energy become highly vulnerable to numerical errors in such a regime. As time proceeds, a strong shock appears in the outer region. Since a strong magnetic field is imposed in the $x$ direction, the plasma can hardly move in the $y$ direction owing to the strong magnetic tension force. On the other hand, the plasma can move freely in the $x$ direction, suggesting that the high beta region elongates in the $x$ direction. These qualitative interpretations are consistent with the simulation results. The spatial profiles of $\rho$ and $\beta$ agree with the results reported in \citet{iijima2021}. 

\subsubsection{Gyrotropic limit}\label{chap5:blast_problem:gyrotropic_limit}

\begin{figure}
	\centering
	\includegraphics[width=\textwidth]{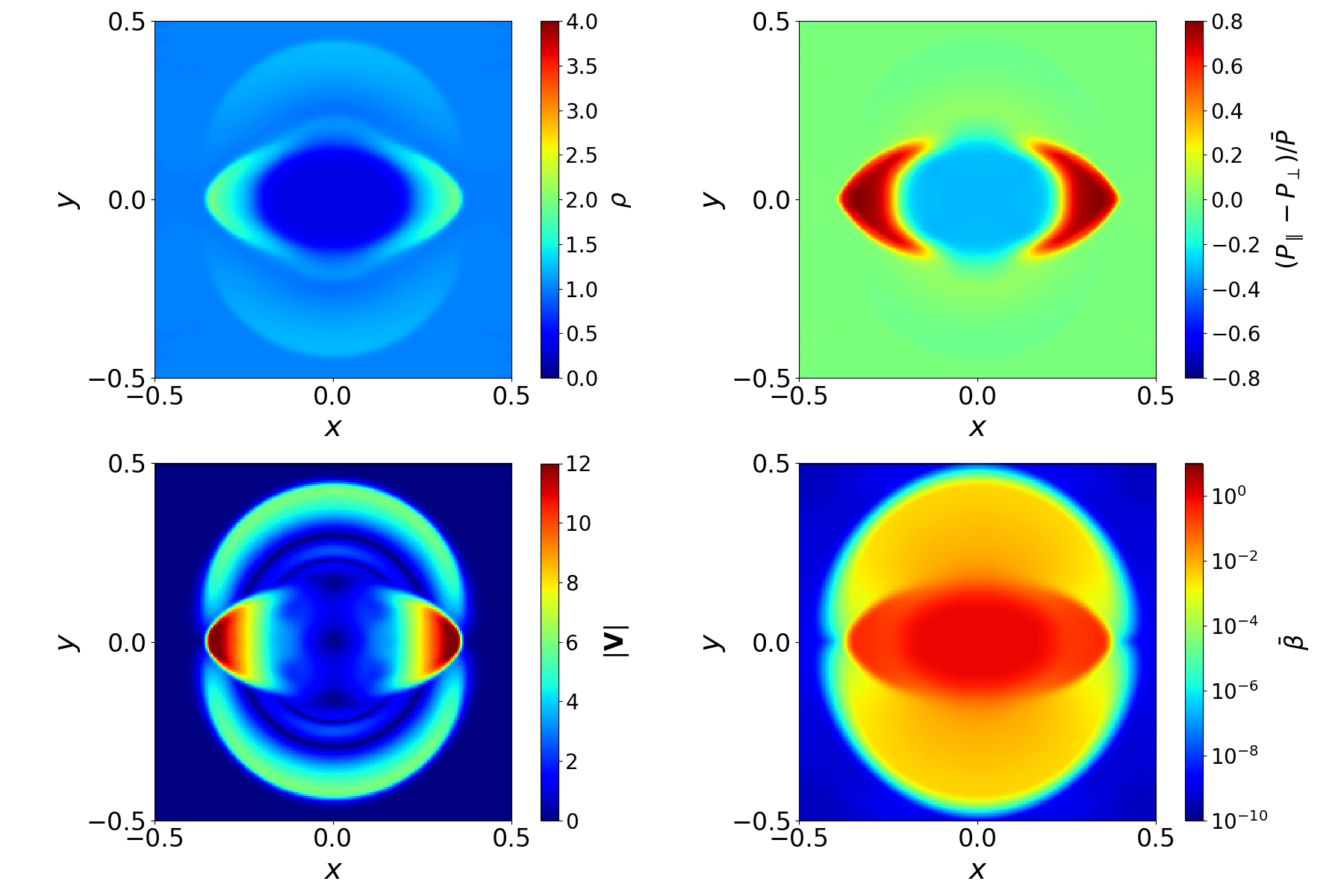}
	\caption{Snapshot of the blast problem in extremely low beta system in the gyrotropic limit at $t = 0.0097$. $\rho, (P_\parallel - P_\perp) / \bar{P}, |\bm{V}|, \bar{\beta}$ are shown. $\bar{\beta}$ is defined as $2 \bar{P} / |\bm{B}|^2$ where $\bar{P} := (P_\parallel + 2 P_\perp) / 3$. A pair of strong jets propagates along the $x$ direction, and the central core also elongates along the $x$ direction. A large $P_\parallel$ appears around the jets, while $P_\perp$ becomes dominant around the core.}
	\label{chap5:fig:blast_problem:gyrotropic_list}
\end{figure}

Figure \ref{chap5:fig:blast_problem:gyrotropic_list} shows snapshots of $\rho$, $(P_\parallel - P_\perp) / \bar{P}$, $|\bm{V}|$, and $\bar{\beta}$ in the gyrotropic limit using the second-order scheme at $t = 0.0097$. Only the second-order result is presented here due to the same reason in the isotropic limit case. Compared with the isotropic limit result, $P_\perp$ dominates near the center of the simulation domain, while $P_\parallel$ dominates in the outer region. As in the isotropic limit, the plasma expands preferentially along the $x$ direction while its motion in the $y$ direction is strongly suppressed by the magnetic tension force. The resulting pressure anisotropy is consistent with this constrained expansion: $P_\parallel$ dominates in the expanding regions along the $x$ direction, whereas $P_\perp$ dominates near the center, where transverse motion is strongly inhibited. This qualitative interpretation is consistent with the simulation results. The spatial profiles agree qualitatively with those reported in \citet{bhoriya2024}, even though that study used a background plasma beta approximately $10^6$ times larger than ours. This qualitative agreement supports the ability of the proposed scheme, including its treatment of numerical heating, to capture pressure-anisotropy effects in extremely low beta plasmas. 


\subsection{Magnetic reconnection}\label{chap5:magnetic_reconnection}

The seventh test simulation is magnetic reconnection with localized resistivity, designed to produce a Petschek-type configuration \citep{petschek1965}. The initial conditions are given by
\begin{align*}
    &\rho = \rho_0 \cosh^{-2}(y / \delta) + \rho_b, \\
    &\bm{V} = (0, 0, 0), \\ 
    &\bm{B} = (0, B_0 \tanh(y / \delta), 0), \\ 
    &P_{ij} = (p_0 \cosh^{-2}(y / \delta) + p_b) \delta_{ij},
\end{align*}
where we set $\rho_0 = 1$, $B_0 = 1$, and $p_0 = B_0^2 / 2 = 0.5$. The background density and pressure are set to $\rho_b = 1.0$ and $p_b = 0.1$, respectively, corresponding to a background plasma beta of $\beta = 0.2$. The current sheet thickness is set to $\delta = 1$. The simulation domain has dimensions of $100\delta \times 50\delta$ and is resolved with $N_x \times N_y = 1000 \times 500$ grid points. The resulting grid spacing is $\Delta x = \Delta y = 0.1\delta$, so that the initial current sheet is resolved with 10 grid spacings across $\delta$. Localized resistivity is prescribed at the center of the simulation box. A resistive term is introduced in the induction equation, taking the same discretization as the nonlinear filtering flux (i.e., transforming the $-\nabla \times (\eta \bm{J})$ term into the form $\delta_{1, \alpha} (...)$). The corresponding Joule heating is added to the diagonal components of the pressure tensor equation, assuming equipartition. We use a sixth-order scheme with a nonlinear filtering flux computed using WENO5-JS reconstruction. Free boundary conditions are imposed in the $x$ and $y$ directions. 

\subsubsection{Isotropic limit}\label{chap5:petschek:isotropic_limit}

\begin{figure}
	\centering
	\includegraphics[width=\textwidth]{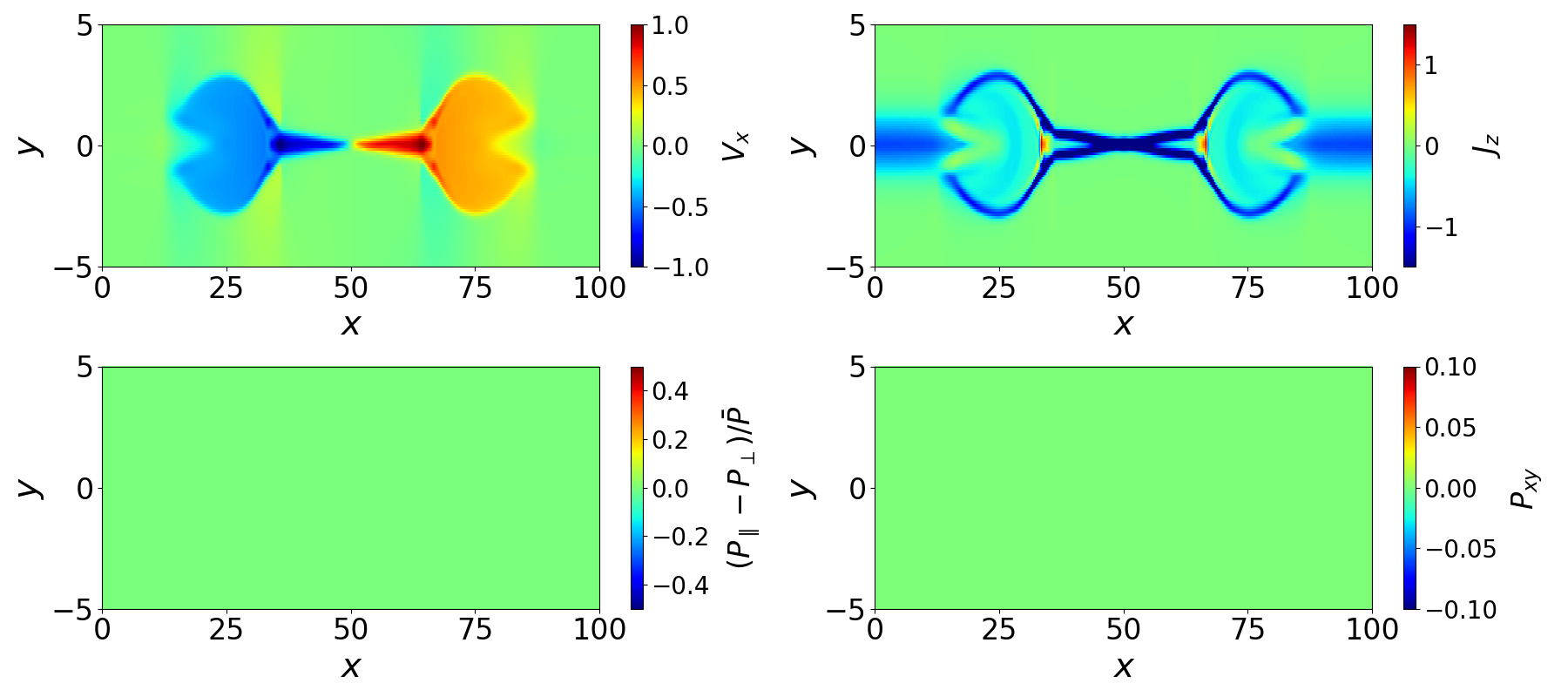}
    \includegraphics[width=0.8\textwidth]{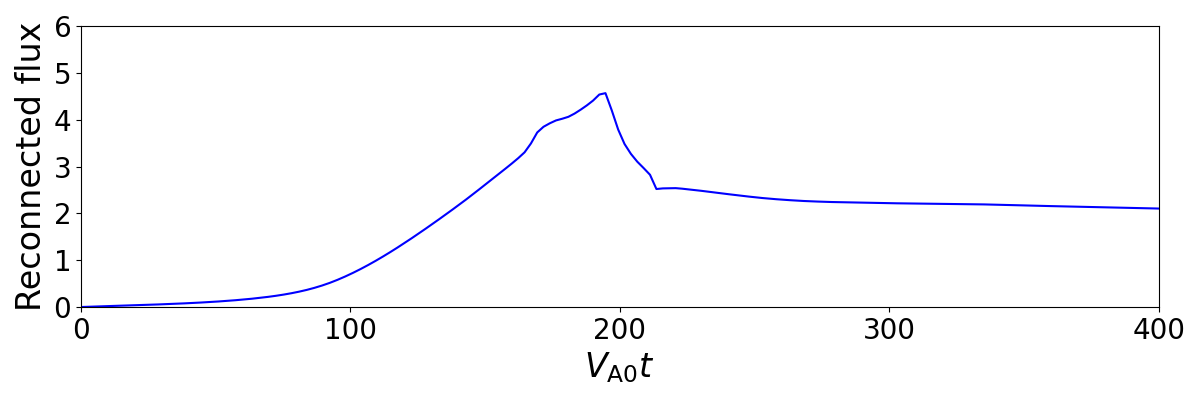}
	\caption{The upper four panels show the snapshots of Petschek reconnection in the isotropic limit at $t = 145.4$. $V_x, J_z, (P_\parallel - P_\perp) / \bar{P}, P_{xy}$ are shown. Here, $\bar{P}$ is defined as $(P_\parallel + 2 P_\perp) / 3$. Reconnection outflows form along the $x$ direction with crab-claw-like structures at the edge. Strong $J_z$ peaks appear between the inflow and outflow regions. The lower panel shows the time evolution of the reconnected flux. Here, the reconnected flux is defined as $\int_{0}^{L_x} |B_y (x, y=0)| \dd x$, where $L_x$ is the simulation box size in the $x$ direction. The reconnected flux increases with time, then suddenly decreases, and finally becomes nearly constant.}
	\label{chap5:fig:petschek_reconnection:isotropic_limit}
\end{figure}

The upper panels of Figure \ref{chap5:fig:petschek_reconnection:isotropic_limit} show snapshots of $V_x$, $J_z$, $(P_\parallel-P_\perp)/\bar{P}$, and $P_{xy}$ in the isotropic limit at $t = 145.4$. Here, $\bar{P}$ is defined as $(P_\parallel + 2 P_\perp) / 3$. Bidirectional reconnection outflows develop from the central diffusion region, with $V_x<0$ on the left and $V_x>0$ on the right. The outflow regions are bounded by narrow current layers, visible as localized enhancements in $J_z$. These current layers correspond to the switch-off slow shocks characteristic of Petschek reconnection \citep{petschek1965,ugai1995,zenitani2011petschek}. Meanwhile, both $(P_\parallel-P_\perp)/\bar{P}$ and $P_{xy}$ remain nearly zero throughout the simulation domain, confirming that the pressure tensor is maintained in an isotropic state. These results show that, in the isotropic limit, the proposed scheme reproduces the basic structure of resistive-MHD Petschek reconnection, including a localized diffusion region, bidirectional outflows, and slow-shock structures. 

The lower panel of Figure \ref{chap5:fig:petschek_reconnection:isotropic_limit} shows the time evolution of the reconnected flux. Here, the reconnected flux is defined as $\int_{0}^{L_x} |B_y (x, y=0)| \dd x$, where $L_x$ is the simulation box size in the $x$ direction. The reconnected flux increses up to nearly $5$ within $V_\mathrm{A0} t = 100$--$200$, corresponding to the time derivative of the reconnected flux (reconnection rate) about $0.05$, consistent with previous studies. The sudden decreases in the reconnected flux occurs because the reconnection outflow reaches the boundary and exits the simulation domain through the free boundary conditions imposed in the $x$ direction. After the outflow exits the simulation domain, the reconnected flux remains approximately constant. 

\subsubsection{Gyrotropic limit}\label{chap5:petschek:gyrotropic_limit}

\begin{figure}
	\centering
	\includegraphics[width=\textwidth]{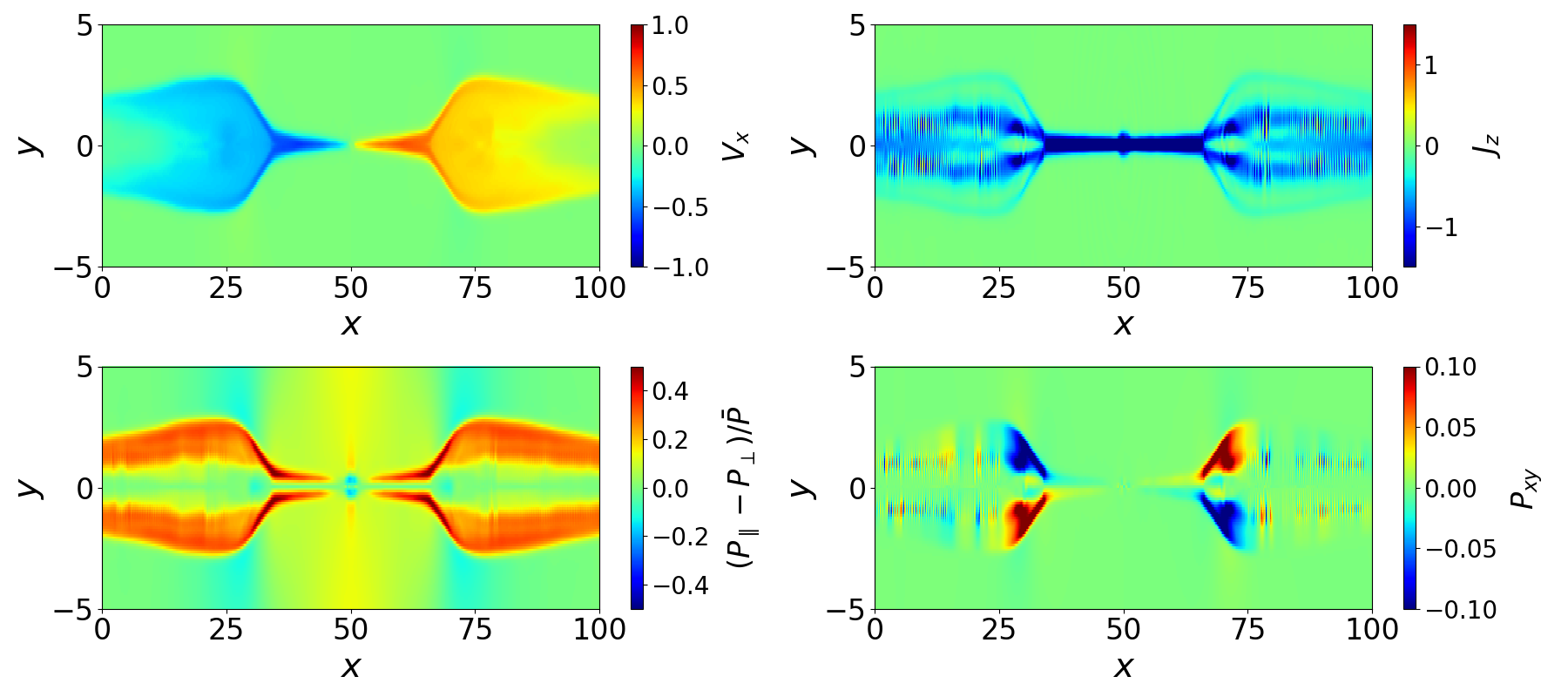}
    \includegraphics[width=0.8\textwidth]{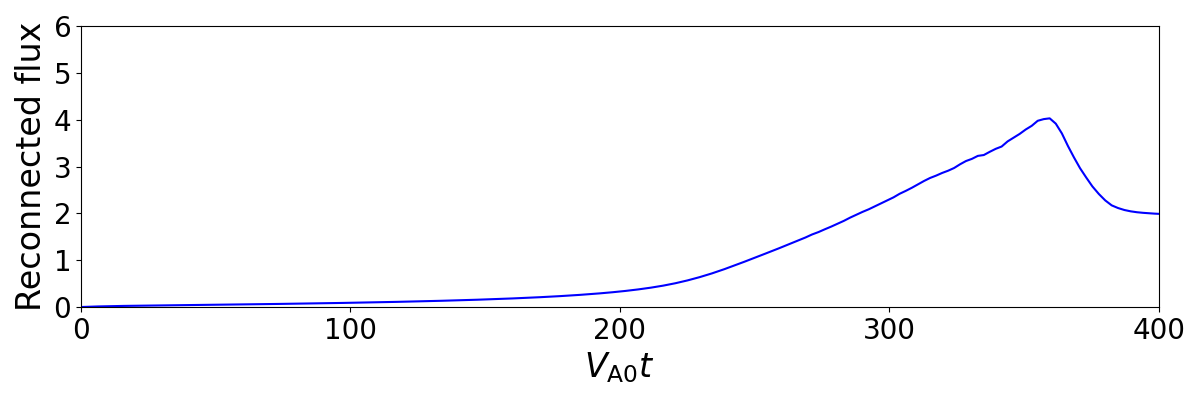}
	\caption{The upper four panels show the snapshots of Petschek reconnection in the gyrotropic limit at $t = 294.7$. $V_x, J_z, (P_\parallel - P_\perp) / \bar{P}, P_{xy}$ are shown. Here, $\bar{P}$ is defined as $(P_\parallel + 2 P_\perp) / 3$. Reconnection outflows form along the $x$ direction. In the outflow region, an elongated current sheet forms, $P_\parallel$ becomes dominant, and a quadrupolar structure of $P_{xy}$ appears. The lower panel shows the time evolution of the reconnected flux. Here, the reconnected flux is defined as $\int_{0}^{L_x} |B_y (x, y=0)| \dd x$, where $L_x$ is the simulation box size in the $x$ direction. The reconnected flux increases with time, then suddenly decreases, and finally becomes nearly constant.}
	\label{chap5:fig:petschek_reconnection:gyrotropic_limit}
\end{figure}

the upper panels of Figure \ref{chap5:fig:petschek_reconnection:gyrotropic_limit} show snapshots of $V_x$, $J_z$, $(P_\parallel - P_\perp) / \bar{P}$, and $P_{xy}$ in the gyrotropic limit at $t = 294.7$. The pair of localized $J_z$ enhancements associated with switch-off slow shocks is absent. Instead, an elongated current sheet develops along the reconnection outflow, indicating that the classical Petschek configuration with standing switch-off slow shocks is not obtained in the gyrotropic limit. This modification of the reconnection structure is consistent with the effects of pressure anisotropy reported in a previous work \citep{egedal2013}. The overall morphology of the reconnection region is qualitatively consistent with that reported in previous CGL-MHD simulation \cite{hirabayashi2013} and PIC simulations \cite{liu2012, fujimoto2016}. Previous studies have shown that switch-off slow-shock solutions can be suppressed when $P_\parallel$ becomes sufficiently larger than $P_\perp$ in the downstream region \cite{karimabadi1995, liu2011theory, hirabayashi2013, akutagawa2026riemann}, which is also consistent with the results shown here. The outflow region also exhibits grid-scale fluctuations that may be associated with the firehose instability, because a strong pressure anisotropy with $P_\parallel>P_\perp$ develops in this region. 

The lower panel of Figure \ref{chap5:fig:petschek_reconnection:gyrotropic_limit} shows the time evolution of the reconnected flux. The reconnected flux increses up to nearly $4$ within $V_\mathrm{A0} t = 200$--$350$, corresponding to the time derivative of the reconnected flux (reconnection rate) about $0.02$--$0.03$. The order of the reconnection rate value is the same as the case in the isotropic limit. The sudden decreases in the reconnected flux occurs due to the same reason in the isotropic limit case. 

\subsubsection{Without isotropization/gyrotropization}\label{chap5:petschek:without_isotopization_and_gyrotropization}

\begin{figure}
	\centering
	\includegraphics[width=\textwidth]{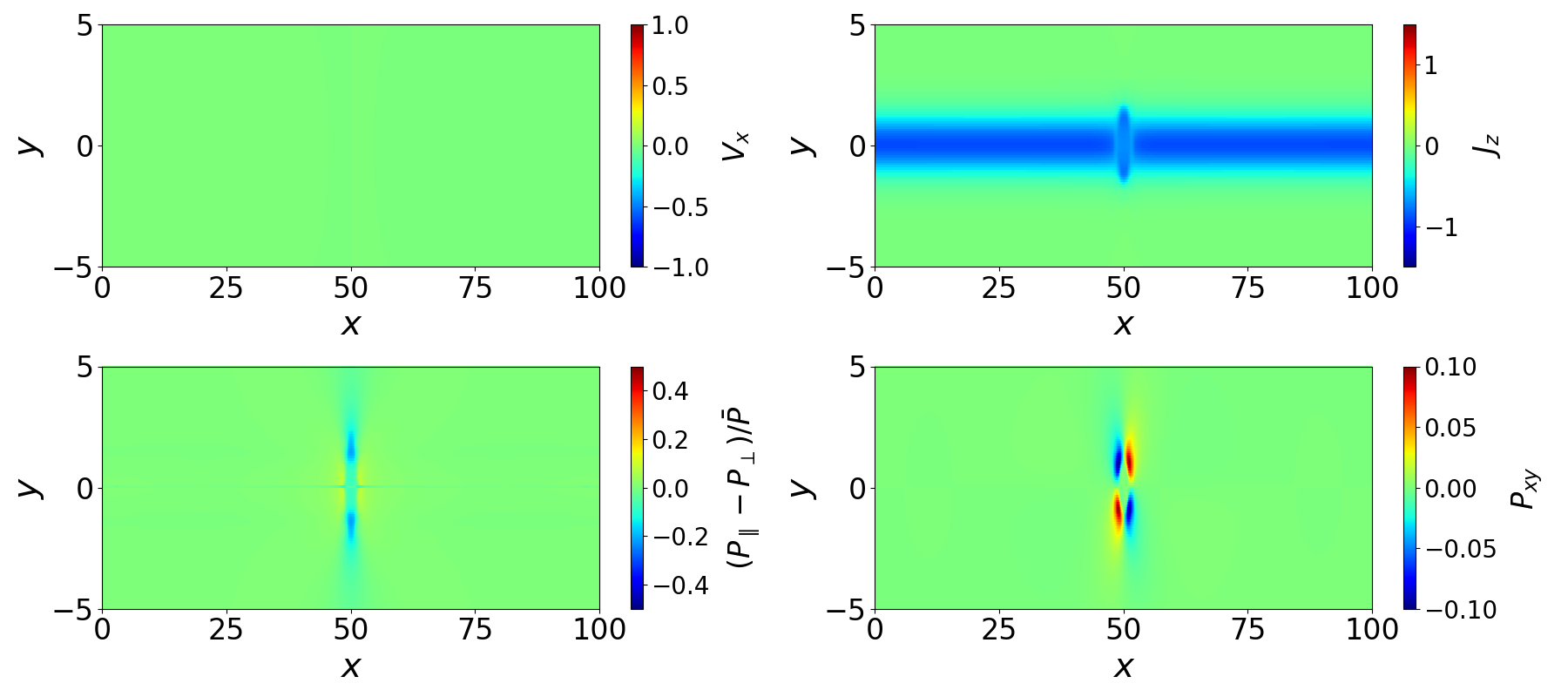}
    \includegraphics[width=0.8\textwidth]{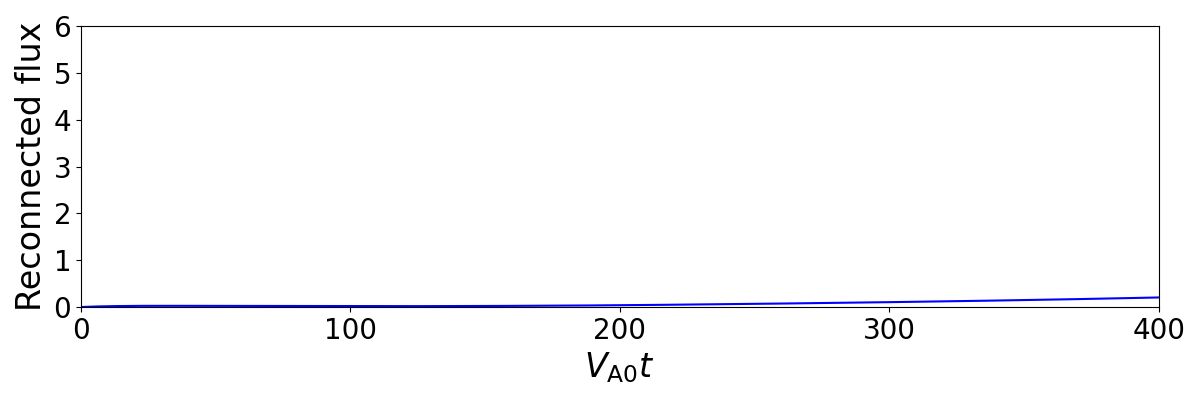}
	\caption{The upper four panels show the snapshots of Petschek reconnection without isotropization/gyrotropization at $t = 377.4$. $V_x, J_z, (P_\parallel - P_\perp) / \bar{P}, P_{xy}$ are shown. Here, $\bar{P}$ is defined as $(P_\parallel + 2 P_\perp) / 3$. No outflow forms. Around the diffusion region, $P_\perp$ becomes dominant, and a quadrupolar structure of $P_{xy}$ can be seen. The lower panel shows the time evolution of the reconnected flux. Here, the reconnected flux is defined as $\int_{0}^{L_x} |B_y (x, y=0)| \dd x$, where $L_x$ is the simulation box size in the $x$ direction. The reconnected flux remains nearly zero throughout the simulation time.}
    \label{chap5:fig:petschek_reconnection:without_isotropization_and_gyrotropization}
\end{figure}

The upper panels of Figure \ref{chap5:fig:petschek_reconnection:without_isotropization_and_gyrotropization} show snapshots of $V_x$, $J_z$, $(P_\parallel-P_\perp)/\bar{P}$, and $P_{xy}$ obtained without isotropization or gyrotropization at $t = 377.4$. In contrast to the isotropic and gyrotropic cases, the current sheet remains close to its initial state, and no appreciable reconnection outflow develops. This result indicates that magnetic reconnection is strongly suppressed when all components of the pressure tensor are allowed to evolve without relaxation. We also find that a localized quadrupolar structure in $P_{xy}$ develops around the central resistive region. In addition, $P_\perp$ becomes dominant over $P_\parallel$ near the diffusion region. These qualitative features are consistent with those reported by \citet{hirabayashi2016} in a one-dimensional Riemann problem for a reconnection layer. 

The lower panel of Figure \ref{chap5:fig:petschek_reconnection:without_isotropization_and_gyrotropization} shows the time evolution of the reconnected flux. The result shows that almost no flux is reconnected within the simulation time, indicating that reconnection does not occur. This is consistent with the spatial profiles in the upper panels of Figure \ref{chap5:fig:petschek_reconnection:without_isotropization_and_gyrotropization}. 


\section{Conclusion}\label{chap6:conclusion}

In this paper, we propose an energy-consistent finite difference scheme for 10-moment MHD that combines the 10-moment MHD model of \citet{hirabayashi2016} with the conventional MHD scheme of \citet{iijima2021}. A nonlinear filtering flux is introduced in the continuity, momentum, and induction equations to suppress numerical oscillations. The numerical heating generated by this filtering flux is added to the diagonal components of the pressure tensor equation so as to preserve total energy conservation. The proposed scheme does not require the eigenmode analysis for 10-moment MHD equations. Moreover, it allows the numerical heating to be partitioned arbitrarily among the diagonal components, whereas this partition is implicitly determined when Riemann solvers are used. These properties make the proposed scheme a flexible and extensible alternative to approaches based on the HLL Riemann solver for 10-moment MHD and to CGL-MHD schemes. 

We performed seven test simulations, summarized as follows:
\begin{enumerate}
    \item \textbf{Circularly polarized Alfv\'en wave propagation} (gyrotropic limit): the L1 convergence is satisfied for all $N_x$ cases when using the second-order scheme, whereas it saturates for the fourth- and sixth-order schemes when $N_x \gtrsim 64$ is used for single wavelength. The total energy conservation error scales with the CFL number in all cases, numerically verifying the energy-consistent design of the proposed scheme. 
    \item \textbf{Parametric decay instability} (isotropic and gyrotropic limits): the linear stage time evolution of the Fourier modes agrees well with the conventional MHD in the isotropic limit case and with the CGL-MHD theory in the gyrotropic limit case. These results suggest that the proposed scheme captures both isotropic and gyrotropic effects. 
    \item \textbf{Shock tube problem} (isotropic and gyrotropic limits, without isotropization/gyrotropization effects): monotonicity is nearly satisfied when using the second-order scheme. Higher-order schemes produce numerical oscillations around the discontinuities, but the nonlinear filtering flux effectively suppresses them. 
    \item \textbf{Orszag--Tang vortex problem} (isotropic limit, with isotropization effect): the fourth- and sixth-order schemes produce smaller-scale structures than the second-order scheme. The $\nabla \cdot \bm{B}$ error can be removed so that the calculation does not crash. Anisotropic effects can be treated qualitatively, particularly the spatial profile of the Sweet-Parker current sheet that forms near the center of the simulation domain.
    \item \textbf{Firehose instability} (gyrotropic limit): the linear stage time evolution of the Fourier mode agrees well with CGL-MHD theory. This result suggests that the proposed scheme can capture anisotropic effects in high beta systems ($\beta \lesssim 10^{10}$).
    \item \textbf{Blast-wave problem} (isotropic and gyrotropic limits): the calculations remain stable in both cases when the background plasma beta is about $10^{-10}$. The result is qualitatively consistent with previous CGL-MHD simulation result \cite{bhoriya2024}. This suggests that the proposed scheme is robust and can capture anisotropic effects in low beta systems ($\beta \gtrsim 10^{-10}$). 
    \item \textbf{Magnetic reconnection} (isotropic and gyrotropic limits, without isotropization/gyrotropization effects): the spatial profiles and reconnection rates in the isotropic and gyrotropic limit cases are qualitatively consistent with previous resistive MHD, CGL-MHD, and PIC simulation results \cite{zenitani2011petschek, hirabayashi2013, liu2012, fujimoto2016}. Magnetic reconnection does not occur without isotropization/gyrotropization effects, which is also consistent with the previous CGL-MHD simulation of a one-dimensional reconnection layer \cite{hirabayashi2013}.
\end{enumerate}
In conclusion, the proposed scheme works well in the isotropic, gyrotropic, and without isotropization/gyrotropization cases. Moreover, it is stable and can capture anisotropic effects across a wide range of plasma beta, from $10^{-10}$ to $10^{10}$. 

The proposed scheme has several limitations. The first is that simulations crash in extremely low beta systems when higher-order schemes are used. We performed simulations of blast-wave problem with $c_\mathrm{diff}$ set to various values ranging from $0.01$ to $10.0$, but the simulations still crash. Developing a more stable scheme for higher-order finite differences is one of the remaining tasks. The second is that we assumed zero heat flux for the 10-moment MHD model. Implementing a more realistic heat flux, such as the closure for Landau damping \cite{hammett1990}, is required to understand kinetic effects in large-scale systems. The third is that we assumed the equipartition of the numerical heating among the three diagonal pressure components without quantitatively verifying this assumption against alternative partitioning strategies. Determining how to partition the numerical heating, and assessing its impact on the pressure tensor evolution, is also one of the remaining tasks. The fourth is introducing the Hall effect, and the gyration effect without using an artificial model; this requires a small timestep, since the ion or electron timescales must be resolved. Developing computationally efficient schemes for 10-moment Hall MHD or 10-moment MHD with an exact gyration effect is also among the remaining tasks. 

Finally, we discuss some applications of the proposed scheme. The first is the physics of the solar wind. The Wind spacecraft and the Parker Solar Probe have reported the existence of temperature anisotropy in the solar wind \cite{bale2009, huang2020psp}. Previous theoretical studies have shown that the parametric decay instability is strongly affected by temperature anisotropy \cite{tenerani2017, saguchi2026}, and this instability is possibly important for solar wind acceleration \cite{shoda2018anis, shoda2018freq, shoda2019}. Understanding the effect of temperature anisotropy on the parametric decay instability, and on models of turbulent coronal heating, is therefore an interesting topic for future work. The second application is the physics of plasmoid-mediated reconnection \cite{loureiro2007, bhattacharjee2009, huang2011, zenitani2020, huang2024, akutagawa2025}. A previous study has reported that temperature anisotropy affects the growth rate of the tearing instability in a Harris current sheet, with the tearing mode growing faster when $P_\perp$ is large \cite{chiou2002}. Moreover, turbulence is known to be affected by anisotropic effects \cite{santos-lima2014}. Investigating the plasmoid instability—which arises from the tearing instability in a Sweet–Parker or reconnecting current sheet and which drives the transition to a turbulent state—is likewise an interesting topic for future work. 


\section*{Acknowledgments}

We thank H. Iijima for helpful discussion about the conventional MHD schemes. This work was supported by JSPS KAKENHI Grant Numbers JP24K00688, JP25K00976, JP25K01052, and by the grant of Joint Research by the National Institutes of Natural Sciences (NINS) (NINS program No OML032402). This work was also supported by JST Sogyo Program (Stage 2), Japan Grant Number JPMJSF2503. Numerical computations were carried out on GPU cluster at the Center for Computational Astrophysics, National Astronomical Observatory of Japan. 



\printcredits

\section*{Data Availability Statement}

The simulation code that support the findings of this study are openly available in GitHub repository \url{https://github.com/keita-akutagawa}.
The simulation results that support the findings of this study are available from the corresponding author upon reasonable request.

\section*{Declaration of competing interest}

The authors declare that they have no known competing financial interests or 
personal relationships that could have appeared to influence the work reported 
in this paper.

\bibliographystyle{elsarticle-num-names}

\bibliography{bib}



\end{document}